\documentclass[%
reprint,
  aps,prab,
  amsmath,amssymb,
  superscriptaddress
]{revtex4-2}
\usepackage{booktabs}
\usepackage{graphicx}
\usepackage{dcolumn}
\usepackage{bm}
\usepackage{hyperref}
\usepackage[mathlines]{lineno}

\newcommand{\IbarakiUniv}{Graduate School of Science and Engineering, Ibaraki University, Mito, Ibaraki 310-8512, Japan}

\begin{document}

\preprint{APS/123-QED}

\title{\textbf{Foundation of Three-Dimensional Spiral Beam Injection\\
Using Canonical Angular Momentum and Symplectic Eigen-Modes
\thanks{Work supported by  JPS KAKENHI Grant Numbers JP19H00673 and JP20H05625.}} 
}%

\author{H.~Iinuma}\email{Contact author: hiromi.iinuma.spin@vc.ibaraki.ac.jp}\affiliation{\IbarakiUniv}

\date{\today}

\begin{abstract}
Aiming for high injection efficiency in three-dimensional spiral injection, the underlying physical principles governing beam formation and matching should be systematically organized within a unified canonical framework. However, a general theoretical framework explaining why particular beam distributions become naturally matched has not yet been established. In this work, a canonical description of three-dimensional spiral injection is developed based on the eigensystem of the symplectic covariance matrix $J\Sigma$.
Canonical modal families are introduced to represent the underlying beam structure, and statistically broadened beam distributions are synthesized around the corresponding modal skeletons while preserving their canonical topology.

Unlike conventional beam-matching methods based on Twiss parameters or eigen-emittance analysis, the proposed framework employs canonical symplectic modes as design variables for beam-family synthesis. It provides a unified description of beam geometry and canonical angular momentum in terms of canonical symplectic modes, and enables beam distributions to be interpreted in terms of their dominant modal structures. Beyond providing a canonical design representation of three-dimensional spiral injection, this approach establishes a direct connection between canonical beam dynamics and experimentally realizable injection beams, thereby providing a theoretical basis for systematic beam synthesis and injection-beam design. The framework further enables the systematic representation, synthesis, and evaluation of statistical distributions of spiral-injection beams in canonical modal space.
\end{abstract}
\maketitle
\section{Introduction}
Previous studies\cite{Iinuma:2016zfu,Iinuma:ipac2025-wepm029, Ogawa:ipac2025-wepm055} have proposed several three-dimensional spiral injection schemes that reported promising injection performance in numerical simulations. 
More recently, the feasibility of three-dimensional spiral beam storage has been experimentally demonstrated~\cite{PRL2026}.
However, their underlying beam dynamics have largely been interpreted from the viewpoint of particle trajectories rather than canonical beam dynamics.
Conventional beam-matching theories describe optical matching and coupled beam transport using Twiss parameters, covariance matrices, and eigen-emittance analysis~\cite{Courant}~\cite{Edwards–Teng},~\cite{LebedevBogaczReview}.
More recently, covariance-matrix formulations and the eigensystem of $J\Sigma$ have been employed for eigen-emittance analysis, normal-mode decomposition, beam diagnostics, and four-dimensional beam characterization~\cite{Duftty},~and~\cite{Groning},~\cite{Xu}.
In particular, covariance-matrix reconstruction has enabled experimental evaluation of four-dimensional beam matrices, eigen-emittances, and interplane coupling~\cite{Xiao}.

However, no general framework has been established for representing three-dimensional spiral beam families as canonical design objects. The purpose of the present work is to formulate such beam families in six-dimensional canonical phase space using the eigensystem of $J\Sigma$.

Three-dimensional spiral injection is intended for compact storage rings with sub-meter orbit radii, where the injection channel and the storage region are connected continuously inside a single solenoidal magnetic field~\cite{Iinuma:2011zz,Iinuma:2016zfu}. Since the beam is transported entirely within external magnetic fields, its dynamics are naturally described in terms of canonical coordinates and canonical momenta.

\begin{figure}[!htb]
   \centering
   \includegraphics*[width=.8\columnwidth]{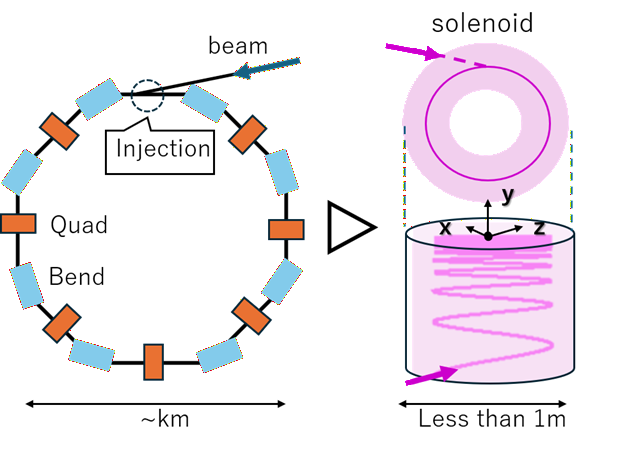}
   \caption{Comparison between conventional beam injection in a large storage ring and three-dimensional spiral beam injection into a compact solenoidal storage region. In the latter, the beam remains immersed in the external solenoidal magnetic field throughout the injection process, motivating the canonical phase-space formulation adopted in this work.
 }
   \label{fig:CompactRing}
\end{figure}

Figure~\ref{fig:CompactRing} illustrates the basic concept of three-dimensional spiral beam injection. In a conventional storage ring, beam transport and storage are realized by a sequence of magnetic elements distributed along the ring circumference. In contrast, three-dimensional spiral injection connects the injection trajectory and the stored orbit continuously inside a compact solenoidal magnetic field. The beam therefore remains immersed in the external magnetic field throughout the injection process. This feature motivates the canonical phase-space description developed in this paper.

Throughout this paper, the origin of the coordinate system is defined at the center of the storage magnet, and a six-dimensional Cartesian canonical phase space is employed. This choice avoids ambiguities associated with the vector potential and provides a globally consistent canonical phase space throughout the entire injection process.
Conventional beam-matching and covariance-matrix-based approaches primarily characterize beam transport or redistribute emittances among coupled degrees of freedom, but they do not provide a design representation for beam families generated in three-dimensional spiral injection.

In the present framework, the covariance matrix $\Sigma$ and the eigensystem of $J\Sigma$ are employed not only to analyze beam distributions but also to define canonical modal families that serve as design variables for beam synthesis.


The remainder of this paper is organized as follows. Chapter~\ref{sec:chap2} introduces the fundamental design problem of three-dimensional spiral injection. This chapter reviews the canonical formalism and the eigensystem of $J\Sigma$. Chapter~\ref{sec:chap3} classifies beam perturbations into single-mode and mixed-mode families. This chapter also develops a modal description of beam families and ribbon reconstruction. Chapter~\ref{sec:chap4} presents a procedure for statistical distribution beam synthesis based on modal beam families. 
Chapter ~\ref{sec:chap4} also presents beam synthesis from canonical modal families.
Chapter~\ref{sec:chap5} summarizes the relationship between the present framework and conventional beam-matching concepts, including BMag, and discusses future extensions.
Finally, the Appendix revisits several beam-design examples previously developed through tracking simulations and reinterprets them within the modal framework introduced in this paper.

\section{Canonical description of a coherent rotating beam}
\label{sec:chap2}
In this section, we introduce the canonical description of the coherent rotating beam underlying three-dimensional spiral injection. 
A preliminary application of the $J\Sigma$ formulation to three-dimensional spiral injection was reported previously in IPAC'26~\cite{Iinuma:ipac2026-thp5620}. The present paper substantially extends that work by introducing canonical modal families, statistical distribution beam synthesis, quantitative modal similarity (later introduced as $C_6$), and the relationship to conventional beam-matching theory.
We first review the kinematic requirements for beam injection into a solenoidal storage magnet and then formulate the corresponding beam dynamics in canonical phase space.
Numerical examples are taken from the demonstration experiment of three-dimensional spiral injection~\cite{PRL2026},~\cite{RehmanThesis},~\cite{Matsushita:ipac2023-mopa118}. 
The formulation itself, however, is general and is equally applicable to the injection designs developed for compact muon storage rings and related projects~\cite{E34PTEP}.

\begin{figure}[!htb]
   \centering
   \includegraphics*[width=.8\columnwidth]{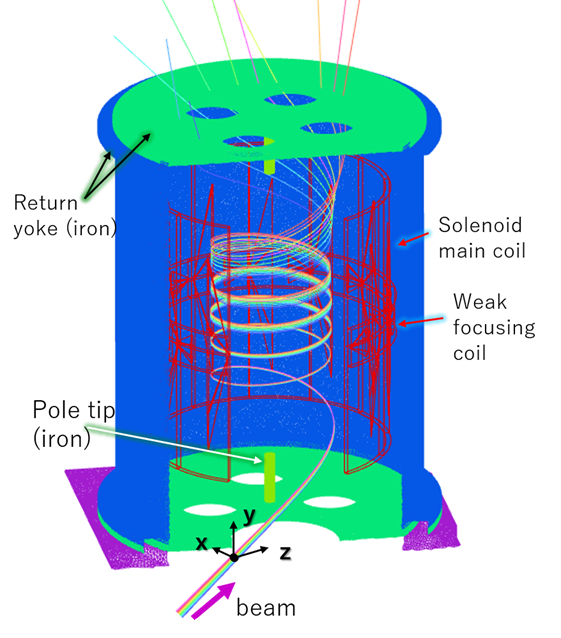}
   \caption{Compact solenoidal storage magnet used in the demonstration experiment. The injection trajectory and the stored orbit coexist within the same magnetic field generated by the solenoid and auxiliary focusing coils, forming a continuous canonical phase space from injection to storage.
}
   \label{fig:SITE}
\end{figure}

Figure~\ref{fig:SITE} shows the compact storage magnet used in the demonstration experiment of three-dimensional spiral injection. The storage region is formed by a solenoidal magnetic field generated by the main coil and shaped by auxiliary magnetic elements, including weak-focusing coils and iron pole structures. Unlike conventional storage rings, the injected beam remains inside the external magnetic field throughout the injection process and gradually approaches the stored orbit through a three-dimensional spiral trajectory.

The beam trajectory shown in Fig.~\ref{fig:SITE} illustrates a representative example of such motion. Because the injection trajectory and the stored orbit coexist within the same magnetic field volume, the beam dynamics are naturally described using canonical phase-space variables. This feature forms the basis of the modal description developed in the following sections.

For particle tracking, mechanical momentum provides a
sufficient description of the particle trajectory and has been
successfully used in previous studies, including the experimental
demonstration of 3-D spiral injection.
However, when beam transport is discussed in terms of transfer matrices, covariance matrices, and phase-space invariants, a canonical phase-space description becomes necessary. The canonical formulation further provides the basis for defining the covariance matrix $\Sigma$ and the eigensystem of $J\Sigma$, which constitute the central framework of this paper.

\begin{figure}[htb]
 \centering
	 \includegraphics*[width=0.6\columnwidth]{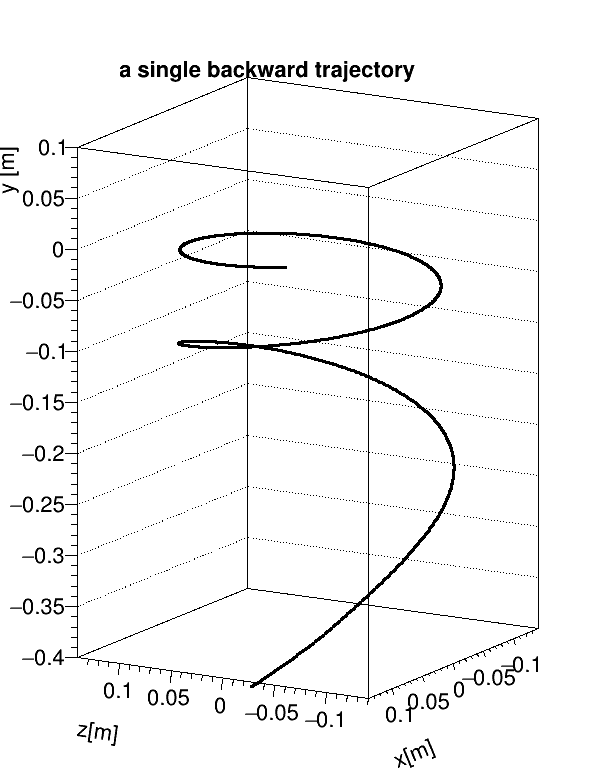}\quad
   \includegraphics*[width=.9\columnwidth]{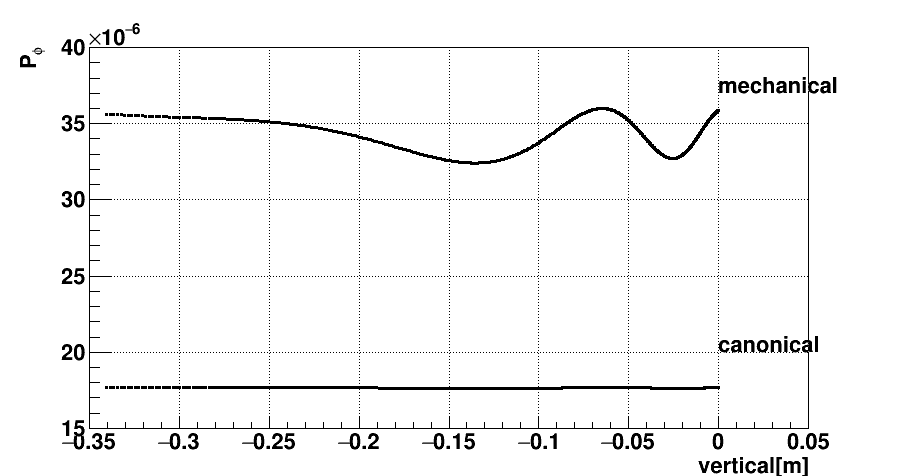}
	 \caption{
     Representative three-dimensional spiral reference trajectory (top) and the corresponding azimuthal angular momentum (bottom). The mechanical angular momentum varies along the trajectory because of the magnetic vector potential, whereas the canonical angular momentum remains nearly conserved. The reference trajectory is used to generate the canonical beam families analyzed throughout this paper.
	 }
      \label{fig:singletrack}
 \end{figure}

Figure~\ref{fig:singletrack} shows a representative three-dimensional spiral trajectory in the storage magnet together with the corresponding azimuthal momentum. The particle gradually approaches the stored orbit while remaining inside the solenoidal magnetic field.

For trajectory calculations, the particle motion can be described using the mechanical momentum. However, the presence of a magnetic vector potential introduces a distinction between mechanical and canonical momenta. As shown in Fig.~\ref{fig:singletrack}, the canonical angular momentum remains nearly constant along the trajectory, whereas the mechanical angular momentum varies significantly as the particle moves through regions of different magnetic field strength.

This observation motivates the use of canonical phase-space variables in the present work. In the following sections, a beam distribution is constructed from rotational copies of a reference trajectory, and the resulting covariance matrix is analyzed through the eigensystem of ($J\Sigma$).

 \subsection{Canonical variables for a track}
Beam dynamics is formulated in canonical coordinates including the vector potential, where the canonical angular momentum plays a central role.
This framework enables a unified treatment of covariance matrices and eigen-emittances.

The phase-space vector of a particle at some point s given by
\begin{equation}
	\vec{q}=(x,~P_x,~y,~P_y,~z,~P_z)^{T},
\end{equation}

 Canonical momenta $\vec{P}$ is obtained as
 \begin{equation}
 P=p+qA
\end{equation}
here, a mechanical momenta $\vec{p}$ and vector potential term $qA$.

Hamiltonian and its differentiation with respect to time are described  as
 \begin{eqnarray}
     H&=&\sqrt{m^2+(P-qA)^2}  \nonumber \\ 
     \dot{q}&=&J\nabla~H
 \end{eqnarray}
 here, $m$ is mass of charged particle and the over dot denotes differentiation with respect to time.   

 Linearization is obtained as
\[
	\delta{\dot{q}}=D_{can}\delta~q
\]
with
 \[
 D_{can}=JH''
\]
where primes denote derivatives with respect to the spatial coordinate.
$J$~is the fundamental symplectic matrix, 
\begin{equation}
    J=diag(J_2,J_2,J2_),~~ J_2=\begin{pmatrix}
   0 & 1 \\
   -1 & 0
 \end{pmatrix}
\end{equation}
and $I$ denotes the $n\times~n$ identity matrix.

The matrix H'' denotes the Hessian matrix of the Hamiltonian,
defined by
\[
H''_{ij}
=
\frac{\partial^2 H}
{\partial q_i \partial q_j}.
\]

Since
\[
H''=(H'')^T
\]
one immediately obtains
\[
{D_{can}}^TJ+JD_{can}=0,
\]
therefore fundamental matrix solution M satisfies \[M^TJM=J.\]
Here \[\dot{M}=D_{can}M.\]

%


Once the particle coordinates are expressed in canonical variables, a beam can be represented as a distribution in a six-dimensional canonical phase space. The statistical properties of such a distribution are described by the covariance matrix $\Sigma$, which will be introduced in the following subsection.

Conventional beam transport and matching methods describe beam distributions primarily in the transverse phase space, while the longitudinal degree of freedom is often incorporated for chromatic or acceptance studies. 
Although such six-dimensional beam representations are widely used in accelerator physics, they are generally not formulated in a fully canonical six-dimensional phase space. Consequently, they do not explicitly preserve the canonical structure required for the present modal formulation.
In the present work, the beam is formulated entirely in canonical phase space using the covariance matrix and the symplectic eigensystem of $J\Sigma$, providing a theoretical framework for understanding coherent rotating beams, modal decomposition, and beam synthesis.

\subsection{Covariance matrix and beam representation}
Once a canonical description has been established for a single particle trajectory, the next step is to represent a beam as an ensemble of particles in the same phase space.

The ideal coherent rotating beam can be regarded as a family of trajectories generated from a single reference trajectory by rotational symmetry about the solenoidal axis. Such a beam therefore possesses an intrinsic modal structure associated with the rotational symmetry of the storage field.

To characterize this ensemble statistically, we introduce the covariance matrix $\Sigma$, which provides a compact representation of the beam distribution in six-dimensional canonical phase space.

\begin{figure}[!htb]
   \centering
   \includegraphics*[width=.9\columnwidth]{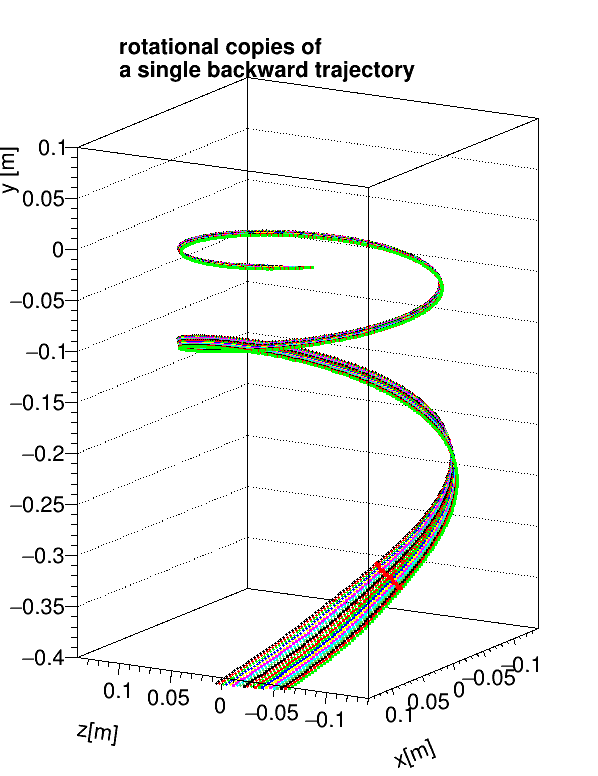}
   \caption{
   Rotational copies of the reference trajectory shown in Fig.~\ref{fig:singletrack}, generated by applying azimuthal rotations around the solenoidal symmetry axis. These trajectories constitute the canonical rotational family and serve as the basis for constructing beam slices and covariance matrices in the following analysis.
  }
   \label{fig:rot-3D}
\end{figure}

To represent a coherent rotating beam, rotational copies of a reference trajectory are generated around the solenoidal axis, as illustrated in Fig.~\ref{fig:rot-3D}. The resulting ensemble forms a family of trajectories sharing the same underlying spiral motion.

At a given longitudinal position, particles are sampled on a transverse slice whose normal vector is defined by the momentum vector of the reference trajectory. The colored points visible near ($y\approx -0.3~\mathrm{m}$) in Fig.~\ref{fig:rot-3D} correspond to such a slice. These sampled particles constitute the beam ensemble used in the present analysis.

\begin{figure}[!htb]
   \centering
   \includegraphics*[width=.9\columnwidth]{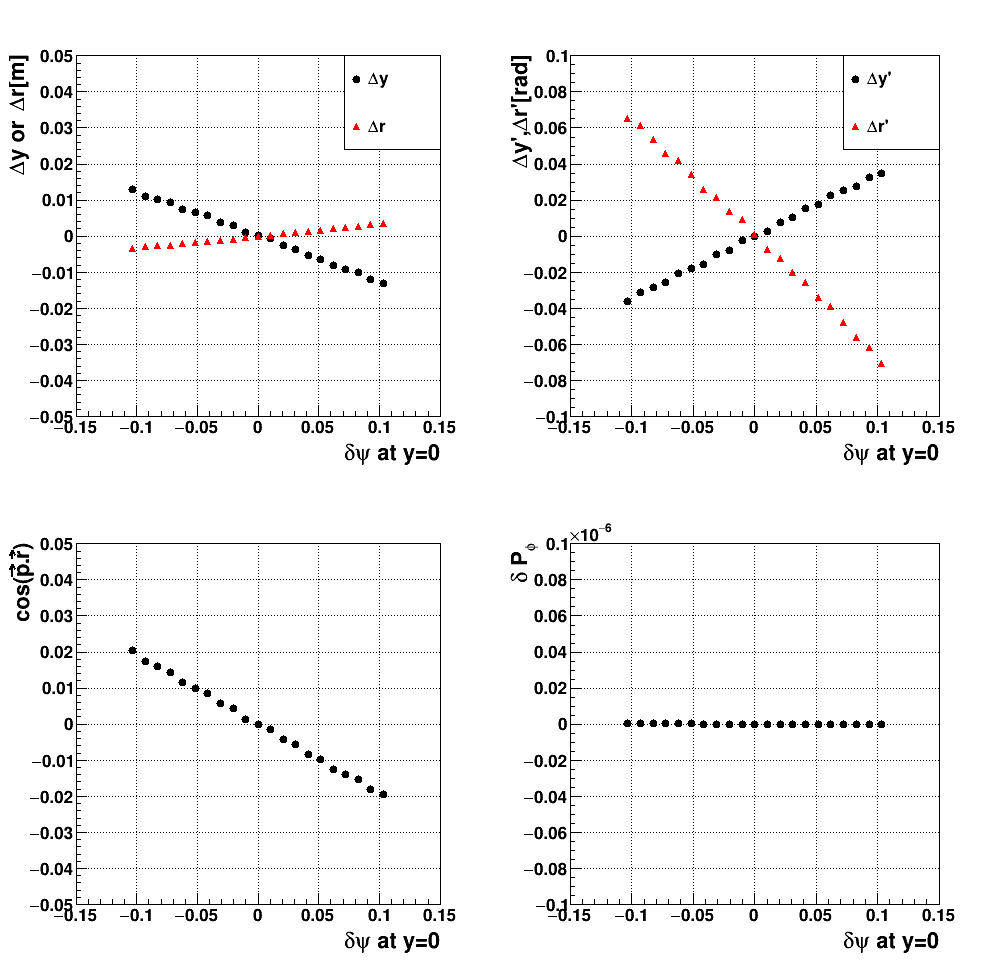}
   \caption{
   Representative beam slice (red markers) obtained from the rotational copies in Fig.~\ref{fig:rot-3D}. The sampled particles define the six-dimensional beam distribution used to construct the covariance matrix $\Sigma$, which is subsequently analyzed through the eigensystem of $J\Sigma$.
   }
   \label{fig:rot-3D_slice_03}
\end{figure}

Let $\vec{q_j}$~denote the canonical phase-space coordinates of the j-th particle. The beam centroid is defined as~$\bar{q}$,
and the phase-space vector of the~$j^{th}$ particle with respect to the beam centroid at slice $i$~is given by
\begin{equation}
	\Delta \vec{q}_{j}=(\Delta~x,\Delta~p_x,\Delta~y,\Delta~p_y,\Delta~z,\Delta~p_z)^{T},
\end{equation}
and the ensemble of~$n$ particles is represented by a~$6\times~n$~matrix $X_{i}$.
The covariance matrix $\Sigma$ at $i$ is then,
\begin{equation}
	\Sigma_{i}=X_{i}X_{i}^{T}.
\end{equation}

In canonical phase space, the evolution of the beam covariance matrix is governed by symplectic transport. If the transfer matrix M satisfies the symplectic condition \[M^T~JM=J,\] 
the covariance matrix evolves according to \[\Sigma_{i+1}=M\Sigma_{i}M^T.\]
This property motivates the use of $\Sigma$ as a beam representation compatible with the underlying Hamiltonian dynamics.

\subsection{The eigensystem of $J\Sigma$}
Although the beam is represented in six-dimensional canonical phase space, its physical interpretation often benefits from a reduced-dimensional description.
A reduced-dimensional description of three-dimensional spiral injection should be regarded as a projection of the underlying six-dimensional canonical phase-space dynamics. The present formulation provides the corresponding symplectic modal structure and defines how beam distributions should be synthesized from it.
Reverse tracking determines a beam that can be injected successfully. The role of the present canonical formulation is not to replace this design process, but to organize the resulting beam into a modal coordinate system suitable for analysis, reconstruction, and optimization.

The covariance matrix $\Sigma$ provides a statistical description of the beam distribution in canonical phase space. However, the covariance matrix alone does not directly reveal the underlying modal structure of the beam. To identify the characteristic modes associated with the beam dynamics, we consider the eigensystem of the matrix $J\Sigma$, where $J$ is the fundamental symplectic matrix.

The eigenvalue problem is defined by

\[
J\Sigma v_k = \lambda_k v_k.
\]
In this paper, we denote $v_k$ as symplectic eigenvector of the $k$-$th$ symplectic eigen-mode.

For a symplectic beam distribution, the eigenvalues appear as complex-conjugate pairs and may be written as

\[
\lambda_k = \pm i\epsilon_k,
\]

where $\epsilon_k$ are the symplectic eigen-emittances.
\begin{eqnarray}
    \epsilon_1&=& 5.59\times 10^{-6} \nonumber \\ 
   \epsilon_2 &=& 7.85\times 10^{-8} \nonumber \\ 
   \epsilon_3 &=& 1.04 \times 10^{-11}
   \label{eq:ribbon_eps2}
\end{eqnarray}

\begin{table}[!hbt]
   \centering
	\caption{ Related vectors obtained from the $J\Sigma$ analysis of Fig.~\ref{fig:rot-3D} and Fig.~\ref{fig:rot-3D_slice_03} }
   \begin{tabular}{l|c}
       \toprule
	     &  Symplectic eigenmode values\\
     & [$\Delta~x$,$\Delta~p_x$,$\Delta~y$,$\Delta~p_y$,$\Delta~z$,$\Delta~p_z$] \\
       \midrule
      Re($v_1$)$\times10^{-2}$ &[-9.74,~9.04,  37.08,  16.66,  89.74,  10.56]   \\ 
       Im($v_1$)$\times10^{-2}$&[ -2.28, -0.094, 0.91, 0.081, 0.00, 0.18] \\
       $\|v_1\|$$\times 10^{-1}$ &[1.00,~~0.90,~~3.71,~~1.66,~~8.97,~~1.06] \\
       \midrule
     Re($v_2$)$\times10^{-2}$& [5.31, -1.71, -56.37, -22.53, -77.80, -8.84]\\
	   Im($v_2$)$\times10^{-2}$&[-3.26, -0.06, 3.52, 0.50, 0.00, 0.11]    \\
       $\|v_2\|$$\times 10^{-1}$ &[0.62,~~0.17,~~5.65,~~2.52,~~7.78,~~0.88] \\
        \midrule
     Re($v_3$)$\times10^{-1}$& [-4.19, -1.10, -6.12, -1.60, -6.35, -0.27]\\
	   Im($v_3$)$\times10^{-2}$&[6.13, 0.56, 6.67, 1.34, 0.00, 1.36]    \\
       $\|v_3\|$$\times 10^{-1}$ &[4.24,~~1.10,~~6.15,~~1.61,~~6.35,~~0.30] \\
       \bottomrule
   \end{tabular}
   \label{tab:eigensystem}
\end{table}

\begin{figure}[!htb]
   \centering
   \includegraphics*[width=.7\columnwidth]{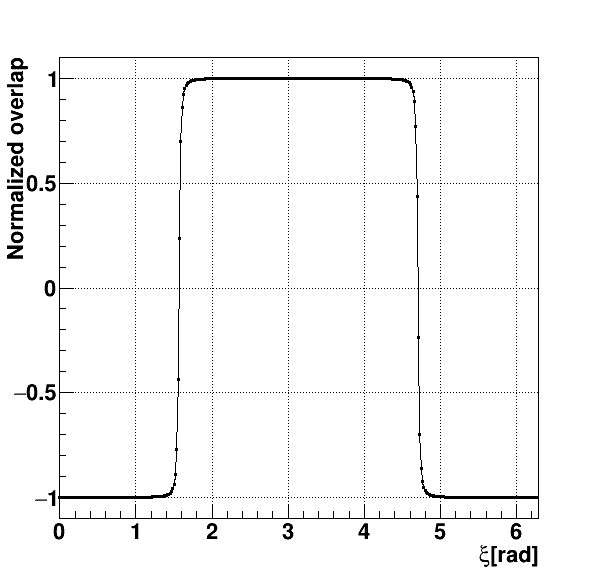}
   \caption{
   Phase dependence of the dominant symplectic eigenvector obtained from the covariance matrix of Fig.~\ref{fig:rot-3D_slice_03}. The eigenvector for k=1 is rotated according to Eq.~\ref{eq:xi}, and the phase is chosen such that the real component becomes dominant. This convention defines the reference canonical mode used throughout the subsequent modal analysis.
   }
   \label{fig:normalizedOverLap}
\end{figure}

Because the symplectic eigenvectors are complex, they are defined only up to an arbitrary phase factor. The real and imaginary parts may therefore be rotated as
\begin{equation}
    v_k(\xi)=Re(v_k)cos\xi+Im(v_k)sin\xi,
    \label{eq:xi}
\end{equation}

without changing the eigenspace. 
In the present work, the phase is fixed such that the real part of the dominant symplectic eigenvector exhibits the maximum overlap with the rotating ribbon. Figure~\ref{fig:normalizedOverLap} shows the normalized overlap between the rotating beam and the phase-rotated eigenvector as a function of $\xi$. The overlap reaches its maximum at $\xi$=0, indicating that the adopted convention provides a natural representation of the rotating mode. The modal coefficients $a_k$ and $b_k$ (or equivalently $c_4$[0–3]) are evaluated using this fixed phase convention throughout this paper.
This choice removes the arbitrary phase degree of freedom while preserving the underlying symplectic eigenspace.

In the present study, particular attention is paid to the dominant eigen-mode associated with the largest eigen-emittance. As shown below, this mode provides a natural description of the coherent rotating beam generated by the rotational symmetry of the solenoidal magnetic field.

To visualize the distribution in each eigen-mode, the deviation vector X is projected onto the modal plane spanned by the real and imaginary parts of the eigenvector. For the $k$-$th$ mode, we define 
\[B_k=[Re(v_k),~Im(v_k)] \]
and determine the modal coordinates 
$(a_k, b_k)$ by a least-squares projection. The residual component $r_k$~represents the part of X not described by this modal plane.
In the numerical implementation, the modal coefficients are stored as a four-component vector $c
_4=(c_4[0],c_4[1],c_4[2],c_4[3])$, where $c_4[0]=a_1$, $c_4[1]=b_1$, $c_4[2]=a_2$, and $c_4[3]=b_2$. 
Throughout this paper, however, the analytical notation ($a_i, b_i$) is adopted for clarity.

\begin{eqnarray}
    X&\simeq& a_k Re(v_k)+b_k Im(v_k)+r_k \nonumber \\ 
    \begin{pmatrix}
        a_k
        b_k
    \end{pmatrix}&=&({B_k}^TB_k)^{-1}{B_k}^TX
    \label{eq:ab}
\end{eqnarray}

\begin{figure}[!htb]
   \centering
   \includegraphics*[width=\columnwidth]{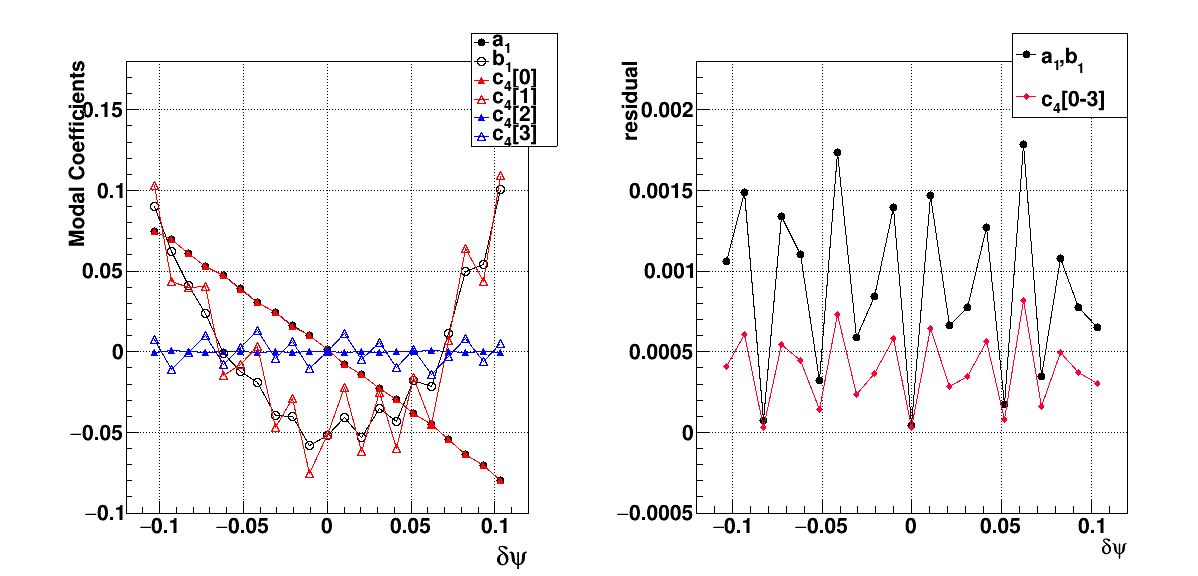}
   \caption{Left:Modal coefficients $c_4[0–3]$ of the rotational copies shown in Fig. 4. The coefficients are obtained by projecting each rotated trajectory onto the first and second canonical eigenmodes listed in Table I according to Eq.~\ref{eq:ab}. The horizontal axis represents the azimuthal rotation angle $\delta\phi$ of the reference trajectory rather than a physical perturbation. This representation forms the basis of the modal-coordinate space shown in Fig.~\ref{fig:rot-3D_modal2}. Right:the residual $r_k$ evaluated according to Eq.~\ref{eq:ab}.}
   \label{fig:rot-3D_modal}
\end{figure}

\begin{figure}[!htb]
   \centering
   \includegraphics*[width=.7\columnwidth]{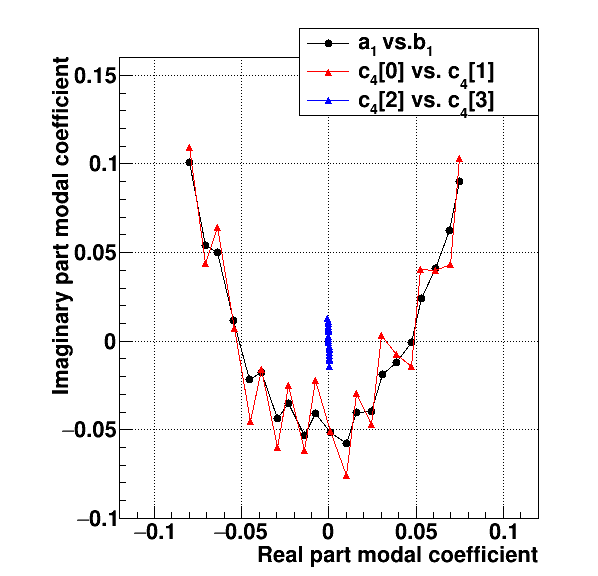}
   \caption{
   Canonical modal-coordinate space constructed from the coefficients $c_4[0–3]$ of Eq.~\ref{eq:ab}. The rotational family forms a simple trajectory in this space, providing the reference representation for the perturbation families analyzed in Chapter~\ref{sec:chap3}.
   Modal-coordinate representation of the rotationally generated beam in the first and second eigen-mode planes of $J\Sigma$. The dominant mode forms a curved one-dimensional family, while the second mode appears as a sub-dominant linear family.
   }
   \label{fig:rot-3D_modal2}
\end{figure}

The modal decomposition is formally constructed using the first two symplectic modes. Figure~\ref{fig:rot-3D_modal2} compares the dominant modal coefficients ($a_1$,$b_1$) with the complete four-component modal representation ($a_1$,~$b_1$,~$a_2$,~$b_2$). Although inclusion of the second symplectic mode further reduces the reconstruction residual, its modal coefficients are found to be significantly smaller than those of the dominant mode for all canonical beam families investigated in this work. The dominant mode alone therefore reproduces the canonical beam skeleton with sufficient accuracy. For simplicity and physical clarity, the following discussion focuses on the dominant modal coefficients ($a_1$,$b_1$), while the second mode is retained only when reconstruction accuracy is explicitly evaluated.




Before introducing perturbed beam families, it is instructive to examine the physical interpretation of the symplectic eigen-modes obtained from the rotationally symmetric beam discussed in the previous section.

\subsection{Geometrical interpretation of the dominant mode}\label{sec:PCA}

As summarized in Table~\ref{tab:eigensystem}, the eigensystem of ($J\Sigma$) exhibits two remarkable features. First, the dominant eigen-emittance is approximately two orders of magnitude larger than the second eigen-emittance, while the third eigen-emittance is nearly zero. This indicates that the rotationally symmetric beam can be described predominantly by a single symplectic mode. Second, the imaginary component of the dominant eigenvector is extremely small compared with its real component. Consequently, the modal plane spanned by the dominant eigenvector is effectively compressed into a one-dimensional curve.

This observation provides an important physical interpretation of the ($J\Sigma$) eigensystem. A conventional PCA analysis identifies only the principal direction of the beam distribution. In contrast, the symplectic eigensystem additionally provides its conjugate direction through the action of the symplectic matrix ($J$). The resulting modal plane forms a natural phase-space representation in which the beam distribution can be expressed by the modal coordinates as shown in Fig.~\ref{fig:R_modal_coff}.


The small imaginary component indicates that the dominant eigenvector is almost real. 
Here, we compare this symplectic structure with conventional Principal Component Analysis (PCA).
PCA identifies only the principal direction of the beam distribution. The small direct overlap between the PCA principal vector and Re($v_1$), together with the large overlap between $Jp_1$ and Re($v_1$), shows that the $J\Sigma$ mode is not merely the PCA direction itself, but its symplectic conjugate representation.
Let $e_1$ and $u_1$~denote the dominant principal component obtained from Principal Component Analysis of the covariance matrix $\Sigma$ and summarize in Table~\ref{tab:PCA}.
\begin{eqnarray}
    e_1&=& 2.2\times 10^{-3} \nonumber \\ 
   e_2 &=& 1.4\times 10^{-6} \nonumber \\ 
   e_3 &=& 9.6 \times 10^{-7}
   \label{eq:ribbon_PCA_e}
\end{eqnarray}
The first three principal components obtained from the PCA of $\Sigma$ are listed in Eq.~\ref{eq:ribbon_PCA_e} for comparison. Since the present distribution represents a skeleton beam, their absolute magnitudes are not physically significant; only their relative magnitudes are considered.
The principal-component vectors corresponding to the three PCA components are summarized in Table~\ref{tab:PCA}.
\begin{table}[!hbt]
   \centering
	\caption{ PCA of $\Sigma$ from Fig.~\ref{fig:rot-3D}. As shown in Eq.~\ref{eq:ribbon_PCA_e}, the primary component is dominant.}
   \begin{tabular}{l|c}
       \toprule
	     &  PCA vectors\\
           & [$\Delta~x$,$\Delta~p_x$,$\Delta~y$,$\Delta~p_y$,$\Delta~z$,$\Delta~p_z$] \\
       \midrule
      $u_1$$\times10^{-2}$ &[~-8.99,~-9.73,~-16.73,~37.24,~-10.55,~89.70]   \\ 
       $u_2$$\times10^{-2}$ &[~18.25,~-64.78,~-17.05,~63.04,~-1.16,~-34.69]   \\
        $u_3$$\times10^{-2}$ &[~20.36,~73.97,~-32.09,~52.37,~7.88,~-16.74]   \\
       \bottomrule
   \end{tabular}
   \label{tab:PCA}
\end{table}

Note that inner products of principle direction $u_1$ from Table~\ref{tab:PCA} and symplectic eigenvector $v_1$ from Table~\ref{tab:eigensystem} are
\begin{eqnarray}
    u_1\cdot Re(v_1)&=&-1.49\times10^{-5} \nonumber \\ 
    Ju_1 \cdot Re(v_1)&=&-0.999
    \label{eq:JRe}
\end{eqnarray}
Therefore, the dominant symplectic mode can be interpreted as the conjugate representation of the principal PCA direction.

The rotationally symmetric beam therefore provides a reference example in which the dominant symplectic mode admits a clear geometrical interpretation. The modal coordinates form a continuous one-parameter family, and the beam distribution can be represented predominantly within a single modal plane.

The comparison demonstrates that PCA captures the dominant geometric variance of the beam distribution, whereas the symplectic eigensystem of $J\Sigma$ uniquely provides the canonical modal pairs and eigen-emittances required for the present beam-family formulation. Therefore, PCA serves as a complementary geometric description, while the $J\Sigma$ eigensystem establishes the canonical modal coordinate system adopted throughout the remainder of this paper.

A natural question is whether this modal structure is preserved when the beam is perturbed away from the ideal rotationally symmetric configuration. To address this question, several representative perturbations are introduced and analyzed in the following sections.

\section{Perturbed Beams and Modal Families}
\label{sec:chap3}
To clarify the intrinsic modal structure of each perturbation family, the following analysis first considers skeleton beams without statistical broadening. This isolates the response of the canonical perturbations before introducing statistically broadened beam distributions in Chapter~\ref{sec:chap4}.

\subsection{Perturbed beam generation}

To investigate the modal response of the beam, four independent perturbations were applied at the vertical position $y=0$ crossing point of the reference trajectory shown in Fig.~\ref{fig:singletrack}. 

\begin{figure}[!htb]
   \centering
   \includegraphics*[width=.9\columnwidth]{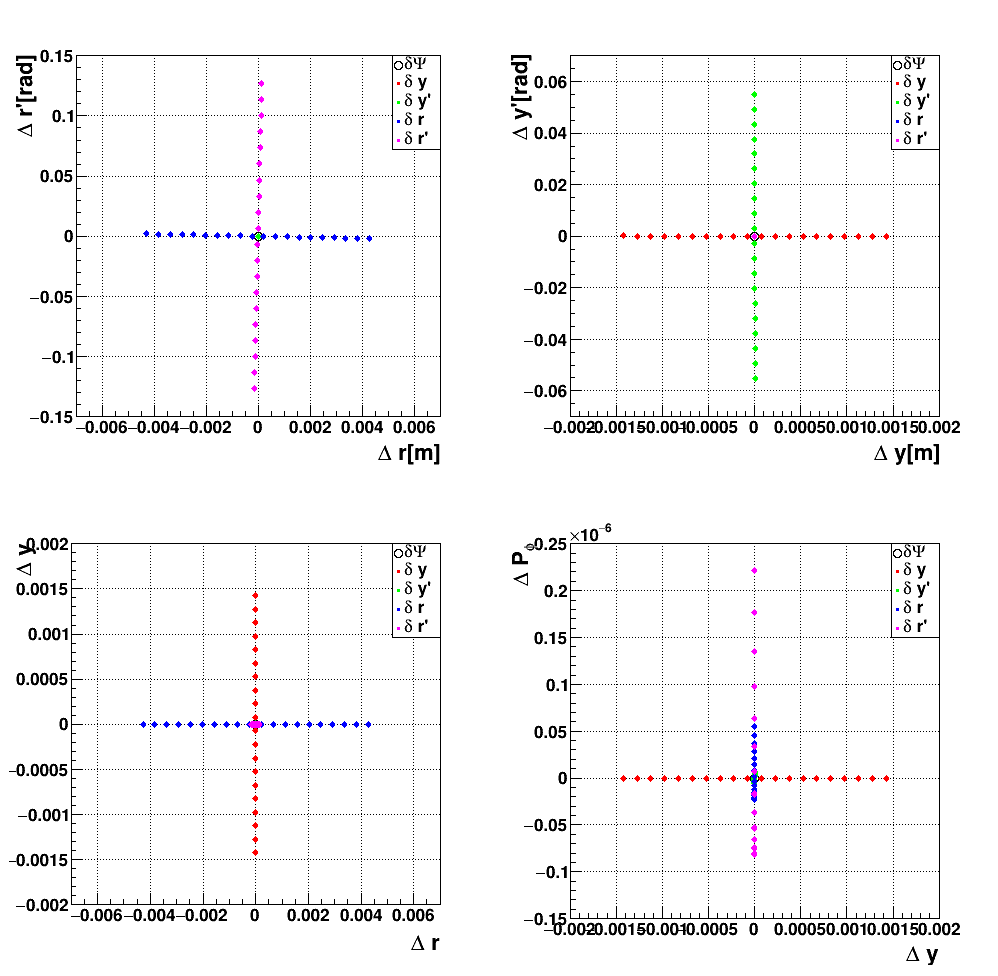}
   \caption{
   Initial perturbation patterns defined at the $y\approx0$ crossing point of the reference trajectory. Four perturbation families are considered: vertical position ($\delta~y$), vertical momentum ($\delta~y'$), radial position ($\delta~r$), and radial momentum ($\delta~r'$), as indicated by the legend. The lower-right panel summarizes the perturbation parameter and the corresponding variation of canonical angular momentum ($\Delta~P_{\phi}$) for each perturbation family. These perturbation patterns serve as the initial conditions for the backward-tracking calculations shown in Figs.~\ref{fig:perturbY} and ~\ref{fig:perturbR}.
Here, the legend identifies the four perturbation families by the lowercase symbols $\delta$, while the corresponding trajectory displacements are denoted by the uppercase symbols $\Delta$ throughout the following discussion.}
   \label{fig:perturbAll}
\end{figure}

These perturbations correspond to offsets in the vertical position ($\delta~y$), vertical momentum ($\delta~y'$) red and green points in Fig.~\ref{fig:perturbAll}. Similarly radial position ($\delta~r$), and radial momentum ($\delta~r'$) in blue and pink points in Fig.~\ref{fig:perturbAll}. 

\begin{figure}[!htb]
   \centering
   \includegraphics*[width=.45\columnwidth]{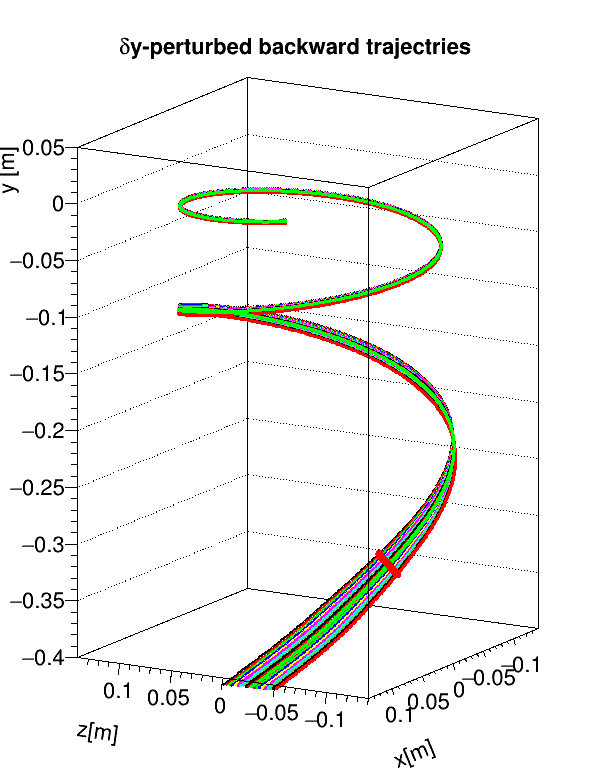} 
   \includegraphics*[width=.45\columnwidth]{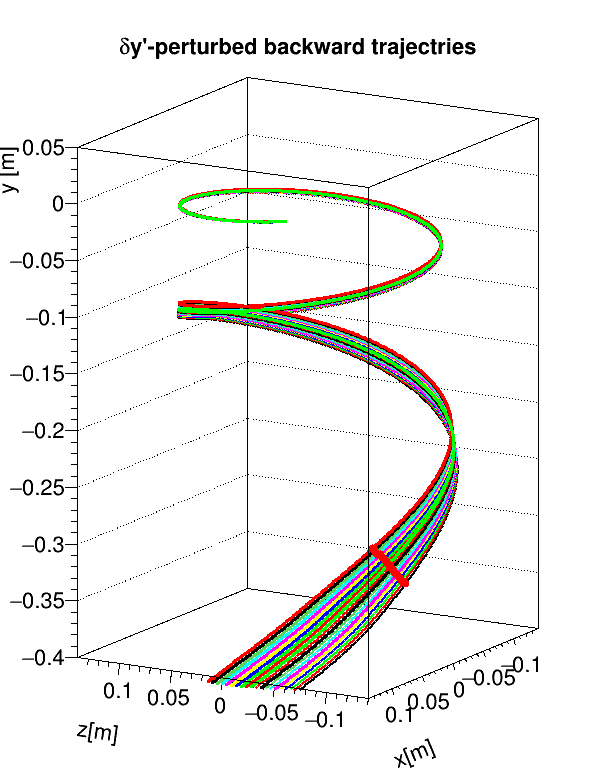}
   \caption{
   Backward-tracked trajectory families generated from the initial perturbation patterns shown in Fig.~\ref{fig:perturbAll} for (left) vertical position perturbations ($\delta~y$) and (right) vertical momentum perturbations ($\delta~y'$). The red markers indicate the beam slice extracted at the reference location ($y\approx~-0.3$~m), corresponding to the same slicing procedure used for the rotational family in Fig.~\ref{fig:rot-3D}. These sliced particle ensembles are used to construct the covariance matrix $\Sigma$.}
   \label{fig:perturbY}
\end{figure}

\begin{figure}[!htb]
   \centering
   \includegraphics*[width=.45\columnwidth]{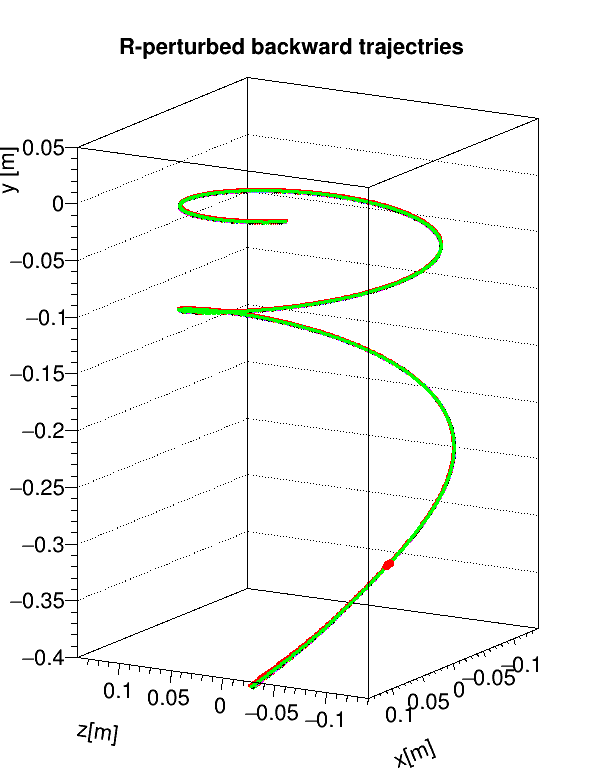} 
   \includegraphics*[width=.45\columnwidth]{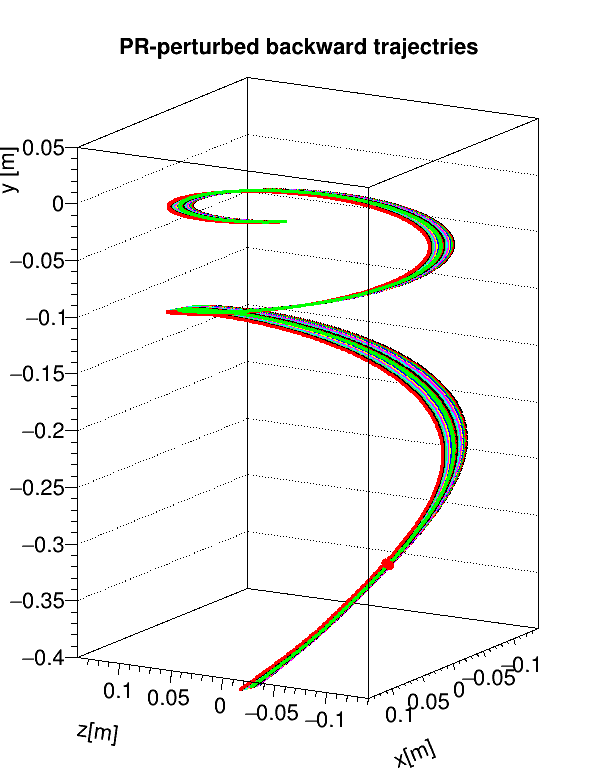}
   \caption{
   Backward-tracked trajectory families generated from the initial perturbation patterns shown in Fig.~\ref{fig:perturbAll} for (left) radial position perturbations ($\delta~r$) and (right) radial momentum perturbations ($\delta~r'$). The red markers indicate the beam slice extracted at the reference location ($y\approx~-0.3$ m), corresponding to the same slicing procedure used for the rotational family in Fig.~\ref{fig:rot-3D}. These sliced particle ensembles are used to construct the covariance matrix $\Sigma$.}
   \label{fig:perturbR}
\end{figure}

 These initial perturbations are subsequently propagated by backward tracking toward the injection point, yielding the trajectory families shown in Figs.~\ref{fig:perturbY} and ~\ref{fig:perturbR}. At the reference location ($y\approx-0.3$~m), the trajectory ensembles are sliced using the same procedure adopted for the rotational family in Chapter II. The resulting particle ensembles are then used to construct the covariance matrix $\Sigma$ for each perturbation family.
Note that the radial perturbations, $\delta~r$ and $\delta~r'$ are defined with respect to the local radial direction of the reference orbit.
Hereafter, lowercase $\delta$ denotes the perturbation type, whereas uppercase $\Delta$ in Figs~\ref{fig:SliceGAt-0}, and as well as the latter figures, denotes the resulting displacement from the reference trajectory.

\begin{figure}[!htb]
   \centering
   \includegraphics*[width=.9\columnwidth]{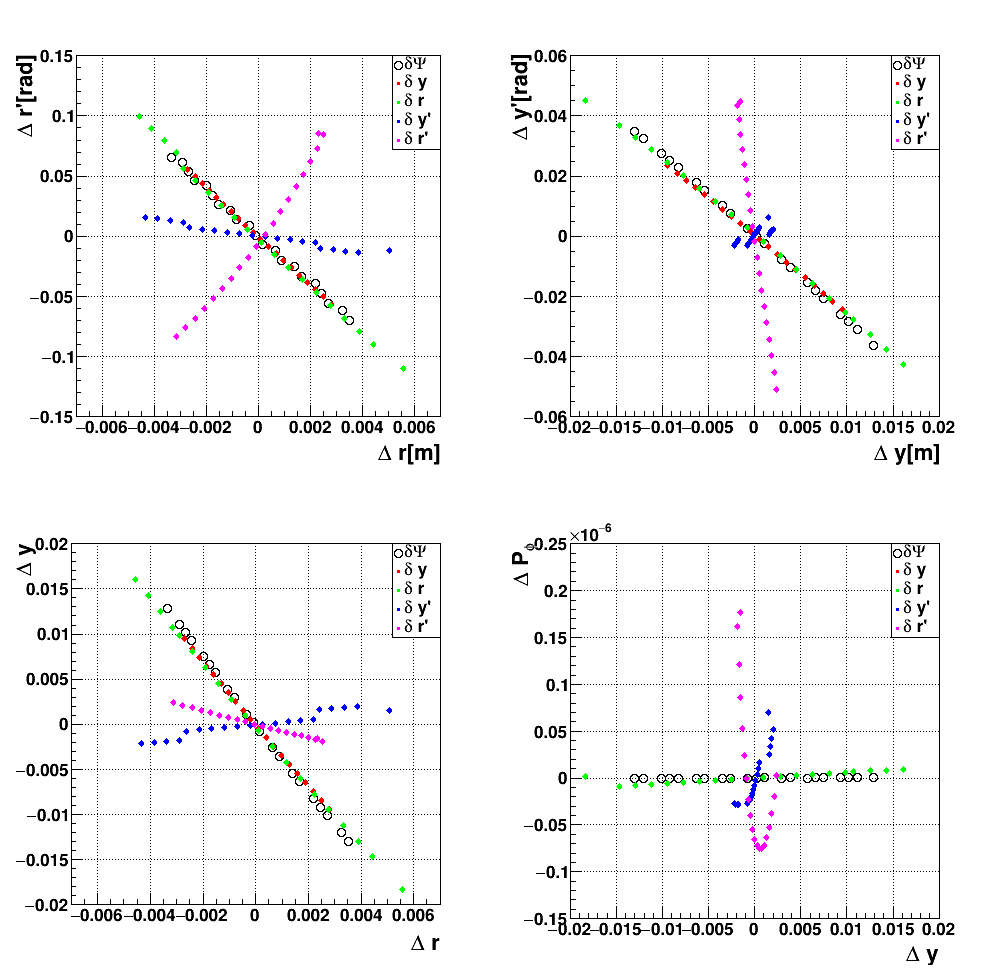}
   \caption{
   Beam slices extracted at the reference location ($y\approx-0.3$~m) from the backward-tracked trajectory families shown in Figs.~\ref{fig:perturbY} and ~\ref{fig:perturbR}. The four panels correspond to the perturbation families $\delta~y$, $\delta~y'$, $\delta~r$, and $\delta~r'$. 
   The lower-right panel summarizes the corresponding canonical angular-momentum response ($\Delta~P{\phi}$) of the sliced beam ensembles. 
These distributions constitute the covariance matrices used for the subsequent symplectic modal analysis.}
   \label{fig:perturbAll-03}
\end{figure}

For each perturbation family shown in Figs.~\ref{fig:perturbAll-03}, a transverse phase-space slice is extracted at the red-marked location. These sliced particle ensembles constitute the input distributions used to construct the covariance matrix $\Sigma$ for each perturbation family. Figures~\ref{fig:perturbAll-03} summarize the corresponding slice distributions together with the canonical angular momentum~($\Delta~P_{\phi}$) evaluated as a function of the applied perturbation.

A clear difference is observed between the perturbation types. 
Vertical perturbations~($\delta~y$~and~$\delta~y'$) leave the canonical angular momentum nearly unchanged, whereas radial perturbations ($\delta~r$ and $\delta~r'$) modify the canonical angular momentum systematically.
This distinction is reflected in the modal structure obtained from the eigensystem of $J\Sigma$ in the following analysis.
Regarding the corresponding canonical angular momentum,shown for each perturbation family, while the vertical perturbations preserve the canonical angular momentum almost unchanged, the radial perturbations produce systematic variations. As shown in the next section, these differences are directly reflected in the eigensystem of $J\Sigma$.

In an ideal uniform solenoidal field, radial perturbations are also expected to preserve a coherent rotational structure. The deviations observed here originate from the realistic magnetic-field distribution of the demonstration experiment, including fringe-field effects and weak-focusing components. As a result, radial perturbations provide a useful example for examining the limitations of the single-mode description introduced in the previous section.

\subsection{Modal decomposition of perturbation families}

The modal coordinates obtained from the dominant symplectic eigenmode are summarized in Table~\ref{tab:puturbMode}\footnote{The first and second symplectic eigenvectors, together with the three symplectic eigenvalues, are provided in Appendix~\ref{sec:SymplecticEigen}.}.

\begin{table}[!hbt]
   \centering
	\caption{ Summary of the first (dominant) symplectic eigenmode of $J\Sigma$ from Fig.~\ref{fig:perturbAll-03}. 
    The second symplectic eigenvectors, together with the three symplectic eigenvalues, are provided in Appendix~\ref{sec:SymplecticEigen}.}
   \begin{tabular}{r|c}
       \toprule
	     &Symplectic Eigenmods' vectors  \\
          & [$\Delta~x$,$\Delta~p_x$,$\Delta~y$,$\Delta~p_y$,$\Delta~z$,$\Delta~p_z$]$\times10^{-2}$ \\
       \midrule
    $\delta~y$  $Re(v_1)$ & [14.74, -7.43, -31.13, -15.22, -91.81, -9.64 ]\\
      $Im(v_1)$ & [1.91, 0.15, -0.97, -0.22,  0.00, -0.06 ] \\ 
 $ \|v_1 \|$ & [  14.87,  7.43,  31.14, 15.22,  91.81,  9.64 ]\\
       \midrule
      $\delta y'$ $Re(v_1)$ & [ -14.25,  7.75,  30.82,  14.84,  91.98,  9.69 ]\\
      $Im(v_1)$& [ -3.58,  0.06,  0.45, -0.19,  0.00,  0.16 ]\\ 
 $\|v_1\|$  &  [  14.69,  7.75,  30.82, 14.84,  91.98,  9.69 ] \\
       \midrule
      $\delta r$~$Re(v_1)$& [ -56.25,  20.17, -23.36,  6.32, -74.80, -15.63]\\
      $Im(v_1)$ &[ -1.88,  0.63, -0.27,  0.27,  0.00, -0.09 ]   \\
      $\|v_1\|$  &  [ 56.28,  20.18, 23.36,  6.32,  74.80, 15.64 ]
\\
\midrule
     $\delta~r'$  $Re(v_1)$ &[  58.24, -2.00,  35.84,  1.99, -72.85,  1.87 ]\\
      $Im(v_1)$& [  1.60, -0.24, -1.08, -0.149,  0.00, -0.34 ]\\ 
    $\|v\|$ &[  58.26,  2.02,  35.86,  1.99,  72.85,  1.90]\\
       \bottomrule
   \end{tabular}
   \label{tab:puturbMode}
\end{table}


For each perturbation family, the trajectory ensemble is projected onto the corresponding modal plane, and the coefficients $(a_1,b_1)$ are evaluated according to Eq.~\ref{eq:ab}.

\begin{figure}[!htb]
   \centering
   \includegraphics*[width=.49\columnwidth]{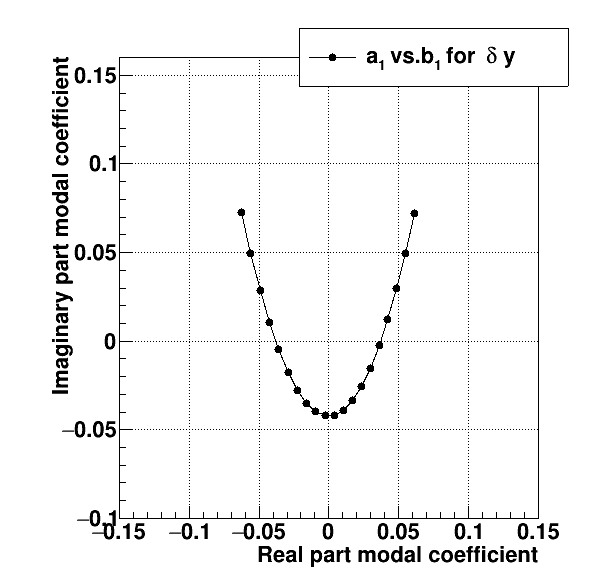} 
   \includegraphics*[width=.49\columnwidth]{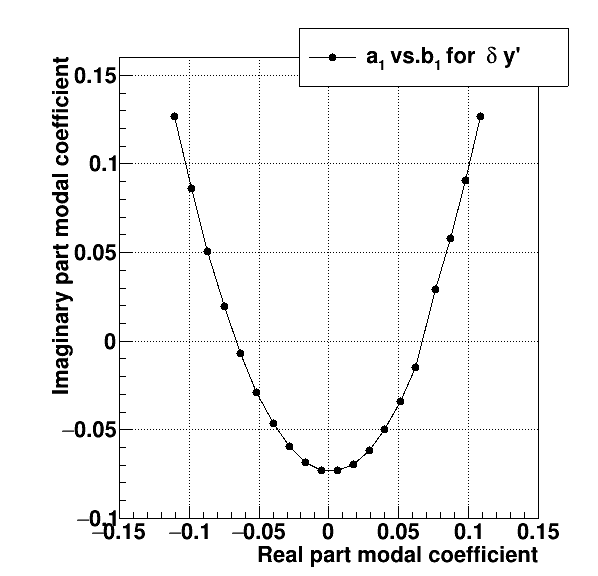}
   \caption{
   Canonical modal skeletons $b_1=f(a_1)$ for the vertical perturbation families: (left) $\delta$y and (right) $\delta$y' defined in Eq.~\ref{eq:ab}.}
   \label{fig:Y_modal_coff}
\end{figure}

\begin{figure}[!htb]
   \centering
   \includegraphics*[width=.49\columnwidth]{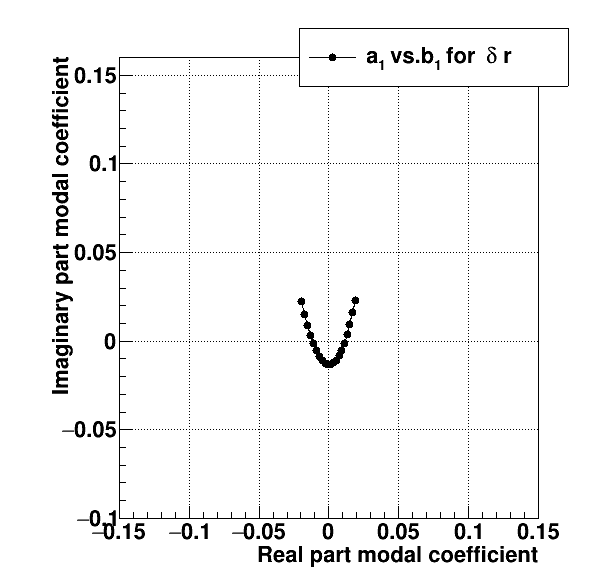} 
   \includegraphics*[width=.49\columnwidth]{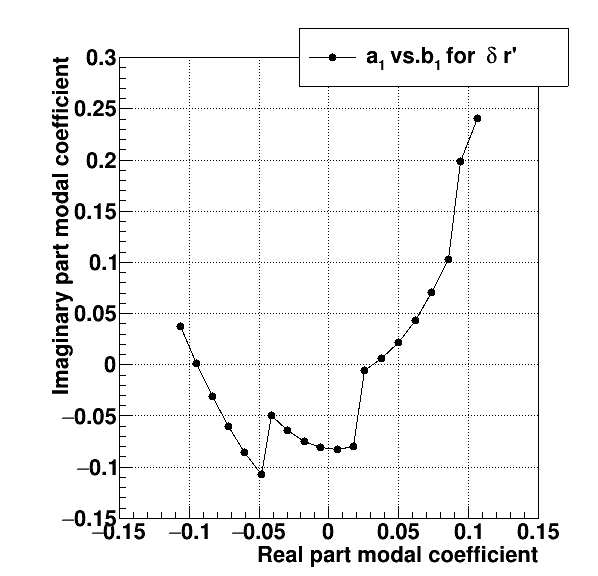}
   \caption{
   Canonical modal skeletons $b_1=f(a_1)$ for the radial perturbation families: (left) $\delta$r and (right) $\delta$r' defined in Eq.~\ref{eq:ab}.   }
   \label{fig:R_modal_coff}
\end{figure}

The resulting relation $b_1=f(a_1)$ defines the canonical modal skeleton. Figures~\ref{fig:Y_modal_coff} and \ref{fig:R_modal_coff} show the skeletons for the vertical and radial perturbation families, respectively. Their distinct shapes characterize the individual perturbation families and provide the reference structures for the statistical broadening introduced in the following section.


The dominant beam structure is governed primarily by $J\mathrm{Re}(v_1)$, while $J\mathrm{Im}(v_1)$ provides an additional component within the modal plane. Although typically small for the skeleton beams considered here, the $J\mathrm{Im}(v_1)$ component retains information on the symplectic modal structure that has no direct counterpart in conventional PCA.

\subsection{Quantitative characterization of canonical beam families}
\label{sec:eta_c6}
\subsubsection{Modal Reconstruction and Projection Residual
}
Although the modal coefficients provide a qualitative classification of the perturbation families, quantitative measures are desirable to evaluate both the validity of the single-mode approximation and the similarity between different canonical beam families.

The quality of the single-mode approximation is quantified by
\begin{equation}
    \eta_1=\frac{|r_{1,i}|}{|X_i|}.
    \label{eq:eta1}
\end{equation}
The corresponding residual vector for the main mode is defined by
\begin{equation}
    r_1=X_i-\left(a_1 \mathrm{Re}(v_1)+b_1 \mathrm{Im}(v_1)\right),
\end{equation}
and note that above equation is same as Eq.~\ref{eq:ab} but set $k=1$.
Here, $X_i$ denotes the $i$-th instance of the same six-dimensional phase-space vector $X$ introduced in Eq.~\ref{eq:ab}, and $r_{1,i}$~denotes its residual from the dominant modal plane.

The parameter $\eta_{1,i}$~quantifies the normalized projection residual of the $i$-th phase-space point with respect to the dominant symplectic modal plane. Smaller values of $\eta_{1,i}$ therefore indicate that the corresponding point is well represented by the modal coordinates ($a_1,~b_1$). 
For each beam family, the characteristic projection residual $\eta_1$~is evaluated from the ensemble and summarized in Table~\ref{tab:eta1}.

\begin{table}[!hbt]
   \centering
	\caption{ Comparisons of $\eta_1$ from Eq.~\ref{eq:eta1} }
   \begin{tabular}{lccccc}
       \toprule
	     modes &  $\delta \psi$ & $\delta~y$ & $\delta~y'$ & $\delta~r$ & $\delta~r'$\\
       \midrule
       $\eta_1\times10^{-2}$& $0.022$ & $0.012$ &$1.0$ &$0.021$&$2.6$ \\ 
       \bottomrule
   \end{tabular}
   \label{tab:eta1}
\end{table}

\subsubsection{Canonical Modal Similarity}
While $\eta_1$ assesses the consistency of the dominant modal reconstruction, we next introduce $C_6$ to quantitatively compare the symplectic eigenvectors obtained from the perturbation families with the corresponding eigenvector of the rotationally symmetric reference beam introduced in Sec.~\ref{sec:chap2}.
\begin{equation}
C_6=\frac{A^{\dagger}B}{|A||B|},
\label{eq:c6}
\end{equation}
where $A=v_{\Psi}$ is the dominant symplectic eigenvector of the rotationally symmetric reference beam, and B=$v_{mode}$ is the corresponding dominant symplectic eigenvector of each perturbation family. The quantity $C_6$ measures the similarity between the dominant symplectic eigenvectors of two canonical beam families and therefore quantifies their modal similarity.

The quantitative comparison is summarized in Table~\ref{tab:C6}.
\begin{table}[!hbt]
   \centering
	\caption{ Comparisons of $C_6$ from Eq.~\ref{eq:c6}.}
   \begin{tabular}{lccccc}
       \toprule
	     modes &  $rot$ & $\delta~y$ & $\delta~y'$ & $\delta~r$ & $\delta~r'$\\
       \midrule
	     $C_6 $   &1.00 &0.997 & 0.997&0.691&0.575\\
       \bottomrule
   \end{tabular}
   \label{tab:C6}
\end{table}

Vertical perturbations ($\delta~y$ and $\delta~y'$)
remain close to the rotational beam
both in reconstruction quality and modal similarity,
whereas radial perturbations ($\delta r$ and $\delta r'$)
show larger deviations. This is a natural consequence of the y-axis being chosen as the solenoid axis, for which vertical perturbations preserve the canonical angular momentum to first order.

\begin{figure}[!htb]
   \centering  
   \includegraphics*[width=0.9\columnwidth]{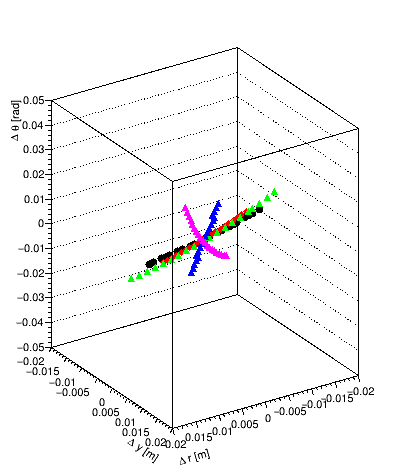}
   \caption{Geometrical representation of the dominant symplectic eigenvectors for the rotational beam and the four perturbation families. The six-dimensional eigenvectors are projected onto the reduced ($\Delta~r,\Delta~y,\Delta~P$) space for visualization. Compared with the phase-space slices in Fig.~\ref{fig:perturbAll-03}, this representation directly illustrates the dominant canonical mode of each perturbation family and facilitates comparison of their geometrical structures.}
   \label{fig:r-y-theta_skeleton}
\end{figure}

\begin{figure}[!htb]
   \centering
   \includegraphics*[width=0.9\columnwidth]{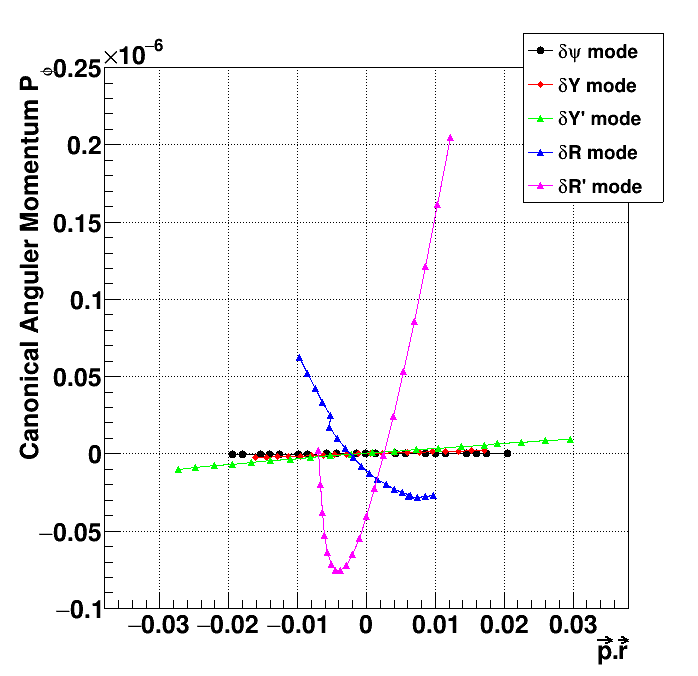}
   \caption{
   Canonical beam families represented as a function of the inner product between the position and canonical momentum vectors. Compared with the displacement-based representation of Fig. 15, the present representation more clearly separates the vertical and radial perturbation families. 
   This scalar representation is also adopted in the following section to compare distributions with different statistical spreads.}
   \label{fig:Proj_r-y-theta_skeleton}
\end{figure}

The canonical modal coordinates introduced in this chapter provide not only a compact description of beam families but also the foundation for constructing reduced-dimensional geometric representations. This canonical modal structure is related to the three-dimensional geometric description traditionally used in spiral beam injection.

\subsubsection{Geometrical Representation in $r$-$y$-$\vec{r}\cdot\vec{P}$ Space}

The quantitative measures $\eta_1$ and $C_6$ introduced above characterize complementary aspects of the canonical beam families in six-dimensional phase space: $\eta_1$ quantifies the projection residual with respect to the dominant modal plane, whereas $C_6$ quantifies the similarity between dominant symplectic modes.
To provide a more intuitive physical interpretation, the same beam families are projected onto the three-dimensional $r$-$y$-$\vec{r}\cdot\vec{P}$ space.

To provide a more intuitive geometrical interpretation of the canonical modal families, the dominant symplectic eigenvectors are projected onto a reduced three-dimensional representation.

Figure~\ref{fig:r-y-theta_skeleton} summarizes the canonical skeletons of the five perturbation families in the reduced ($r,y,\vec{r}\cdot\vec{P}$) space. Although all trajectory families originate from the same reference orbit, they exhibit distinct geometrical structures in the canonical configuration space. Figure~\ref{fig:Proj_r-y-theta_skeleton} further compares the canonical angular momentum as a function of the modal inner product. The vertical perturbation families ($\delta~y$ and $\delta~y'$) preserve the canonical angular momentum almost unchanged, whereas the radial perturbation families ($\delta~r$ and $\delta~r'$) exhibit systematic variations. This difference is a natural consequence of the y-axis being chosen as the solenoid axis.
These observations reveal two distinct classes of canonical beam behavior. This observation motivates the classification of beam families into single-mode and mixed-mode perturbations, which forms the basis of the ribbon-beam construction.
The single-mode canonical skeletons identified here serve as the fundamental design objects for beam synthesis. 



In the following chapter, beam distributions are generated by statistically broadening these canonical modal families, and the resulting canonical structures are examined.

\section{Canonical Modal Beam Synthesis}
\label{sec:chap4}
%

In the previous chapter, each perturbation family was represented by a one-dimensional canonical modal skeleton described by the modal coordinates ($a_k$,$b_k$). The skeleton defines the underlying ribbon structure of the beam and is uniquely determined from the dominant symplectic eigenvectors. 
The present chapter extends this framework to statistical distributions constructed around the canonical skeleton while preserving its modal structure.

Let the modal skeleton be represented by
\begin{equation}
b_0=f(a_0)
\end{equation}
where ($a_0,b_0$) denotes a point on the canonical ribbon. 

The modal skeleton introduced in Chapter~\ref{sec:chap3} defines the centerline of each canonical modal family in the (a,b) plane. 
In the present chapter, statistically broadened beam distributions are generated by introducing statistical spreads around this skeleton.
\begin{figure}[!htb]
   \centering
   \includegraphics*[width=0.52\columnwidth]{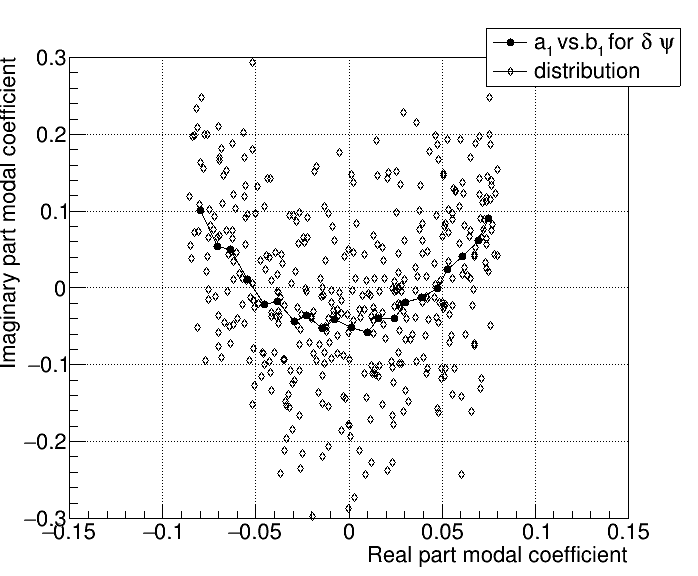} 
   \includegraphics*[width=0.45\columnwidth]{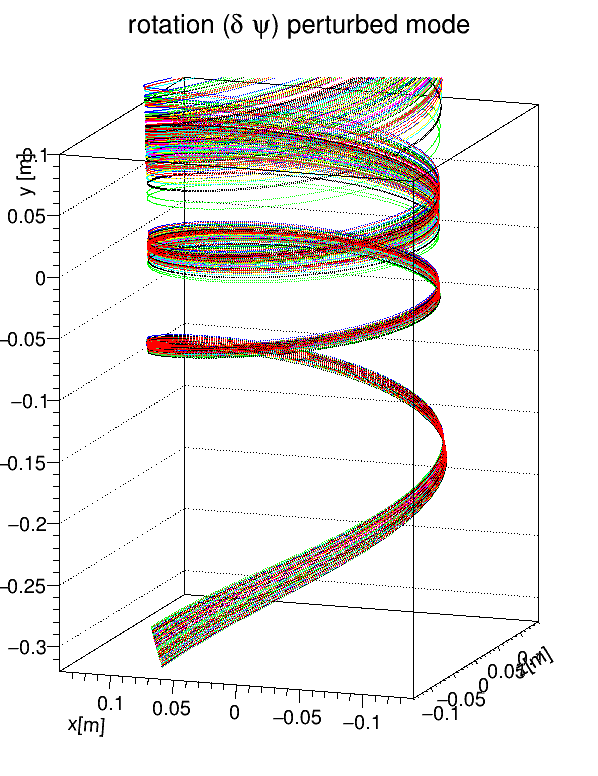}
   \caption{Finite-distribution beam generated as a probability distribution surrounding the canonical modal skeleton by use of Eq.~\ref{eq:gauss}.}
   \label{fig:rot_eps2}
\end{figure}
Left side plot of fig.~\ref{fig:rot_eps2} illustrates an image of the beam-generation procedure using the dominant modal family of the rotating beam as an example. The black curve represents the modal skeleton obtained in Chapter~\ref{sec:chap3}, while the red points denote a statistically broadened distribution generated by adding Gaussian perturbations around the skeleton.

A statistical beam distribution is generated by assigning statistical perturbations
\begin{eqnarray}
  a_1&\rightarrow&a_1+\sigma_{a_1}g_{i}, \nonumber \\  
  b_1&\rightarrow&b_1+\sigma_{b_1}h_{i}  
  \label{eq:gauss}
\end{eqnarray}
where $g_i$ and $h_i$ are independent standard Gaussian random variables, and $\sigma_{a_{1}}$ and $\sigma_{b_1}$ denote the corresponding Gaussian widths. These widths are chosen empirically from the spacing of neighboring skeleton points in order to demonstrate the synthesis procedure.
The width $\sigma_{b_1}$ was chosen empirically so that the vertical beam size near $y~\approx~0$ remained within approximately $\pm~0.01$~m (see the left bottom panel of Fig.~\ref{fig:SliceGAt-0}).
For the statistically broadened distribution, the $J\Sigma$ decomposition yields a first symplectic eigenvalue whose magnitude is larger than that obtained for the corresponding skeleton distribution in Eq.~\ref{eq:ribbon_eps2}. No direct physical correspondence between this change and the imposed statistical spreads is assumed here.

\begin{eqnarray}
    \epsilon_1&=& 1.23\times 10^{-5} \nonumber \\ 
   \epsilon_2 &=& 4.07\times 10^{-9} \nonumber \\ 
   \epsilon_3 &=& 4.89 \times 10^{-12}
   \label{eq:thick_eps}
\end{eqnarray}

\begin{table}[!hbt]
   \centering
	\caption{ Related vectors obtained from the $J\Sigma$ analysis of~Fig.~\ref{fig:rot_eps2}.
    Despite this increase in the first eigen-emittance, the direction of the first eigenvector is found to remain essentially unchanged from Table~\ref{tab:eigensystem}.}
   \begin{tabular}{l|c}
       \toprule
	     &  Symplectic eigenmode values\\
     & [$\Delta~x$,$\Delta~p_x$,$\Delta~y$,$\Delta~p_y$,$\Delta~z$,$\Delta~p_z$] \\
       \midrule
      Re($v_1$)$\times10^{-2}$ &[-9.76,~9.06,  37.60,  16.70,  89.53,  10.50]   \\ 
       Im($v_1$)$\times10^{-2}$&[ -12.50, -0.383, -1.63, -0.083, 0.00, 0.27] \\
       $\|v_1\|$$\times 10^{-1}$ &[0.98,~~0.91,~~3.76,~~1.67,~~8.95,~~1.05] \\
       \midrule
     Re($v_2$)$\times10^{-2}$& [-70.60, -2.06, ~47.64, -4.24, -51.00, -2.68]\\
	   Im($v_2$)$\times10^{-2}$&[~0.00, -1.22, -5.69, -0.82, 8.93, 0.20]    \\
       $\|v_2\|$$\times 10^{-1}$ &[0.62,~~0.17,~~5.65,~~2.52,~~7.78,~~0.88] \\
        \midrule
     Re($v_3$)$\times10^{-1}$& [-6.15, -0.31, 5.04, -0.39, -34.08, -0.50]\\
	   Im($v_3$)$\times10^{-2}$&[0.00, 4.02, 30.33, 4.88, -39.15, -0.55]    \\
       $\|v_3\|$$\times 10^{-1}$ &[6.15,~~0.51,~~5.88,~~0.63,~~5.19,~~0.07] \\
       \bottomrule
   \end{tabular}
   \label{tab:eigensystemRotGauss}
\end{table}


This construction does not modify the underlying modal family itself; instead, it generates a statistical ensemble surrounding the canonical skeleton. 

Consequently, the dominant symplectic eigenvector remains approximately unchanged, indicating that the underlying modal orientation is preserved under the statistical broadening considered here.

The objective of this chapter is therefore not to redefine the modal families introduced in Chapter III, but to investigate how statistical broadening modifies the physical beam distribution while preserving its underlying modal structure. Particular attention is paid to the distribution of canonical angular momentum around the skeleton ribbon and to the statistical spreads associated with each perturbation family.

\subsection{Statistically Broadened Perturbation Families}
\begin{figure}[!htb]
   \centering
   \includegraphics*[width=0.52\columnwidth]{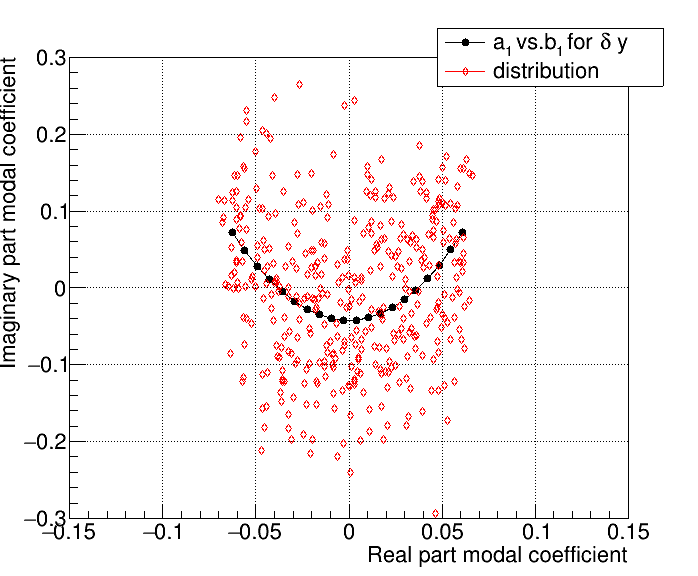} 
   \includegraphics*[width=0.45\columnwidth]{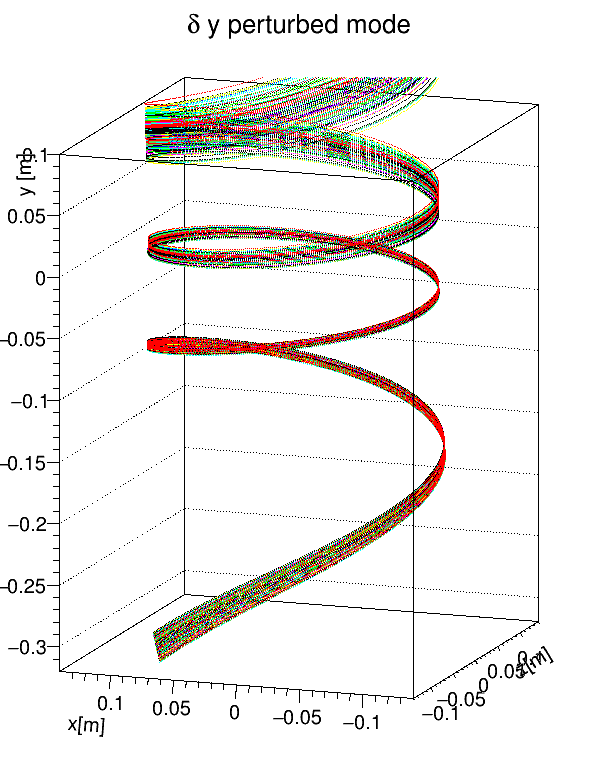}
   \caption{
   Forward-tracked trajectories generated from the finite-distribution beam shown in Fig. 18 for (left) distribution obtained by Eq.~\ref{eq:gauss} and (right) the vertical momentum perturbation ($\delta~y$). The beam distributions preserve the canonical modal topology identified for the corresponding skeleton beams while acquiring a finite phase-space extent through Gaussian broadening. The red markers indicate the beam slices extracted at the $y\approx0$ crossing point for the analysis presented in Fig.~\ref{fig:SliceGAt-0}.}
      \label{fig:Y_a1_b10}
\end{figure}

\begin{figure}[!htb]
   \centering
   \includegraphics*[width=0.52\columnwidth]{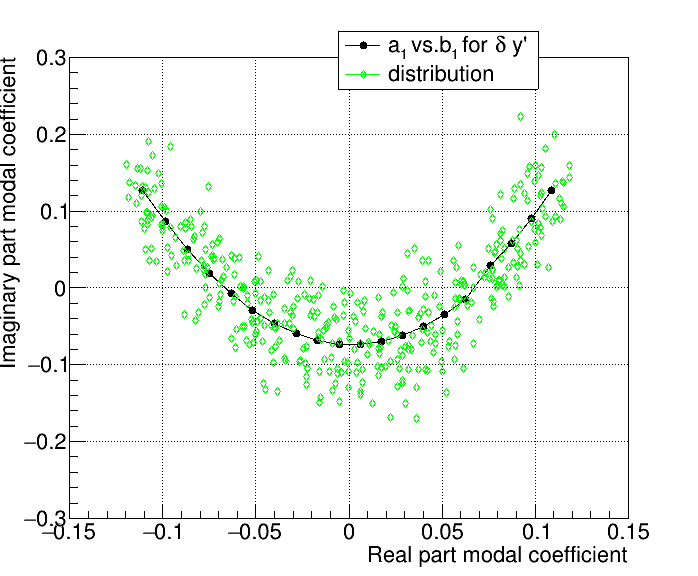} 
   \includegraphics*[width=0.45\columnwidth]{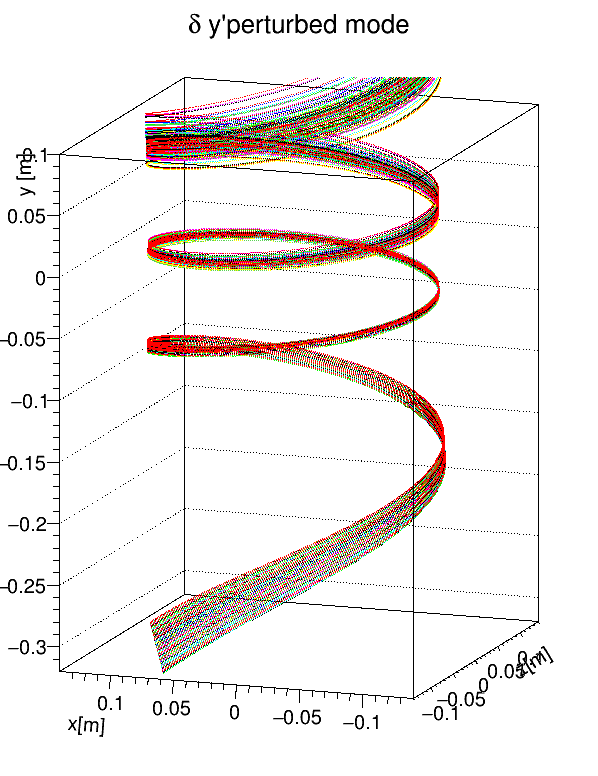}
   \caption{
   Same as Fig~\ref{fig:Y_a1_b10} but for $\delta y'$.}
      \label{fig:PY_a1_b2}
\end{figure}

\begin{figure}[!htb]
   \centering
    \includegraphics*[width=0.52\columnwidth]{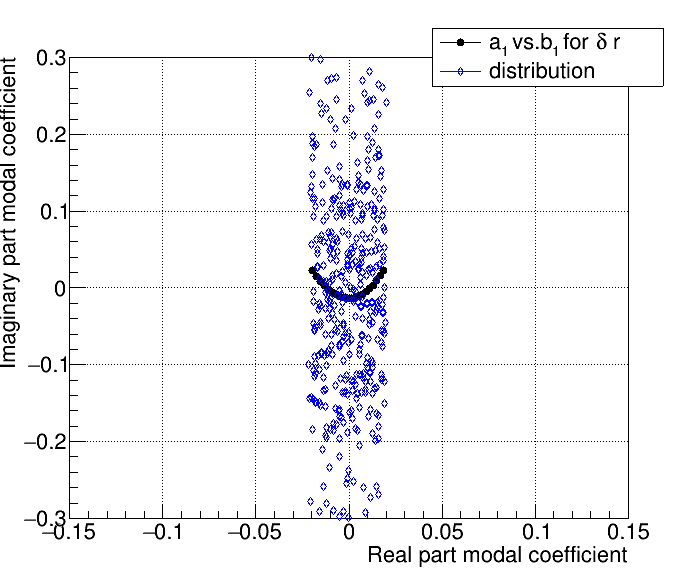}
   \includegraphics*[width=0.45\columnwidth]{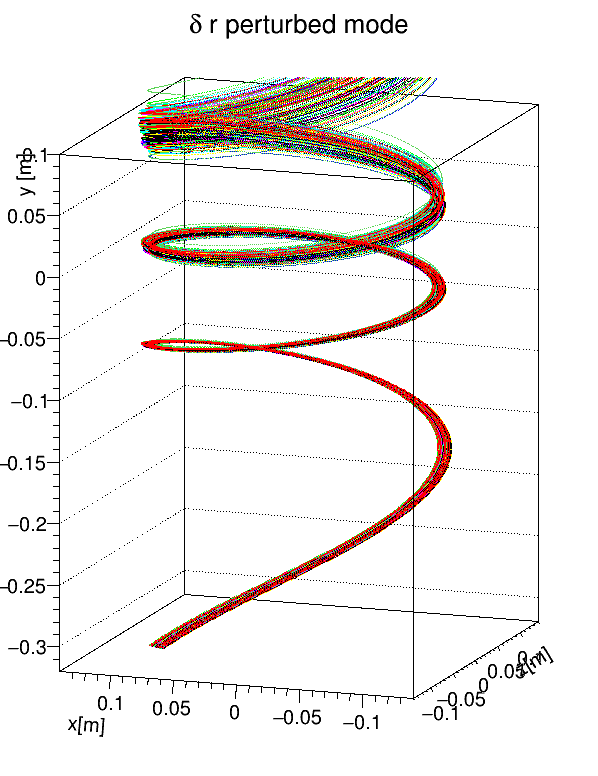} 
   \caption{
   Same as Fig~\ref{fig:Y_a1_b10} but for $\delta r$.}
      \label{fig:R_a1_b10}
   \label{fig:R_a1_b40}
\end{figure}

\begin{figure}[!htb]
   \centering
  \includegraphics*[width=0.52\columnwidth]{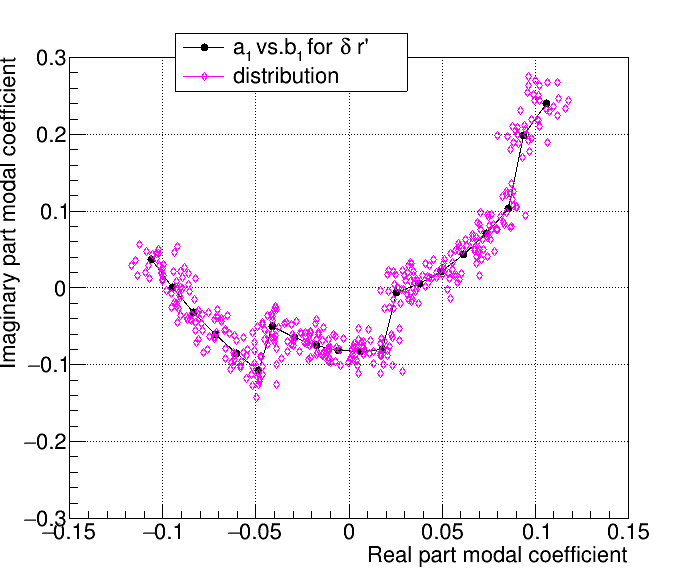}
   \includegraphics*[width=0.45\columnwidth]{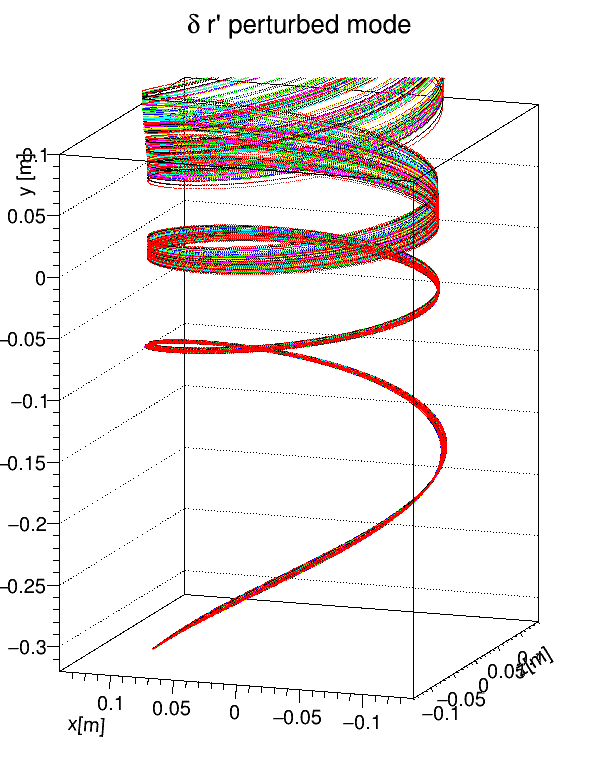}
   \caption{
   }
   \label{fig:PR_a1_b-1}
\end{figure}

The canonical skeletons identified in the previous chapter are used to construct Gaussian broadening beam distributions for each perturbation family. This chapter presents the resulting beam distributions and their canonical properties after particle tracking. Figures~\ref{fig:Y_a1_b10} to ~\ref{fig:PR_a1_b-1} show statistical beam distributions corresponding to the four perturbation families.

The relation (b=f(a)) derived from the symplectic eigenvectors characterizes not only the distribution around a particular reference trajectory, but also a family of reference trajectories sharing the same local modal structure. Within the range shown, the reference trajectory can therefore be chosen at different locations using the common eigenvectors, with the relative distribution described in the same modal basis.

Thus, the common modal basis provides a unified representation of both the relative distribution and the choice of reference trajectory within the examined range.

\begin{figure}[!htb]
   \centering  
    \includegraphics*[width=0.9\columnwidth]{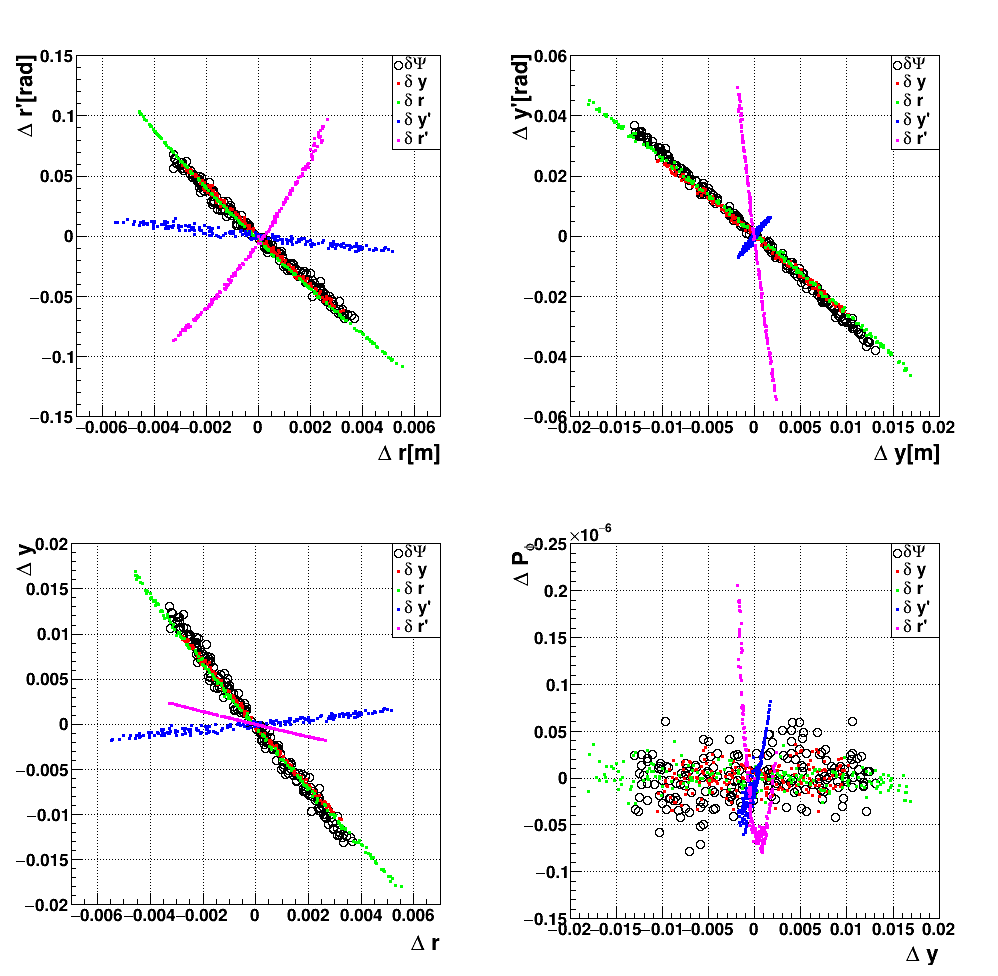}
   \caption{Statistical beam distributions at the injection point ($y\approx-0.3$~m),
   generated by Gaussian broadening around the canonical modal skeletons as in Fig.~\ref{fig:perturbAll-03}. 
   At this stage, the beam distributions preserve the canonical modal families introduced in Chapter~\ref{sec:chap3}, while the canonical-angular-momentum distributions remain only weakly correlated with the vertical displacement.
   These distributions serve as the initial conditions for the forward tracking shown in Figs.~\ref{fig:Y_a1_b10}~and~\ref{fig:PR_a1_b-1}.}
   \label{fig:SliceGAt-3}
\end{figure}


\begin{figure}[!htb]
   \centering  
    \includegraphics*[width=0.9\columnwidth]{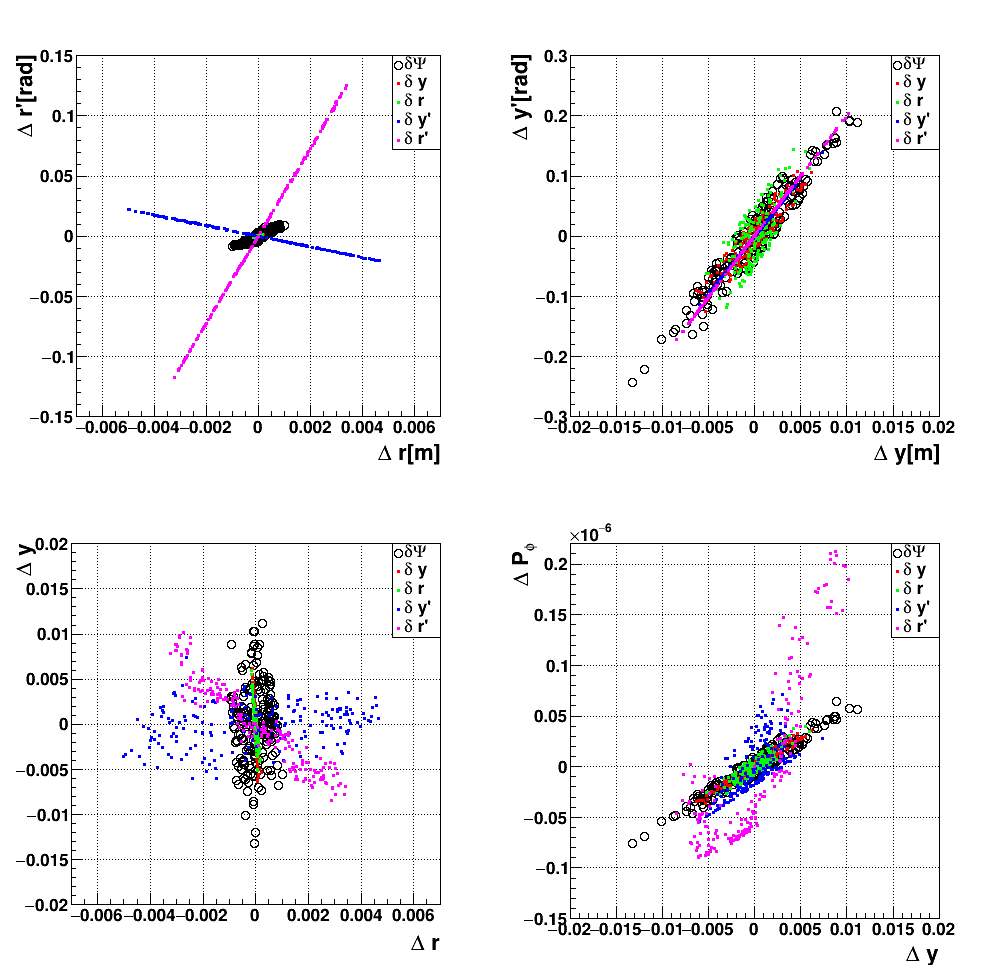}
   \caption{
   Beam slices extracted near the~$y\approx0$~crossing point from the forward-tracked statistical distribution beams shown in Figs.~\ref{fig:Y_a1_b10} and ~\ref{fig:R_a1_b40}. Compared with the initial distributions in Fig.~\ref{fig:rot-3D_slice_03}, the rot-family perturbations ($\delta~\psi$, $\delta~Y$, and $\delta~Y'$) develop a clear correlation between the vertical displacement and the canonical angular momentum, whereas the radial perturbation families remain more broadly distributed. The resulting beam width remains within approximately $\lvert\Delta~y\rvert\leq~0.01$~m, validating the Gaussian broadening adopted in Eq.~\ref{eq:gauss}.}
   \label{fig:SliceGAt-0}
\end{figure}
\begin{figure}[!htb]
   \centering  
    \includegraphics*[width=0.7\columnwidth]{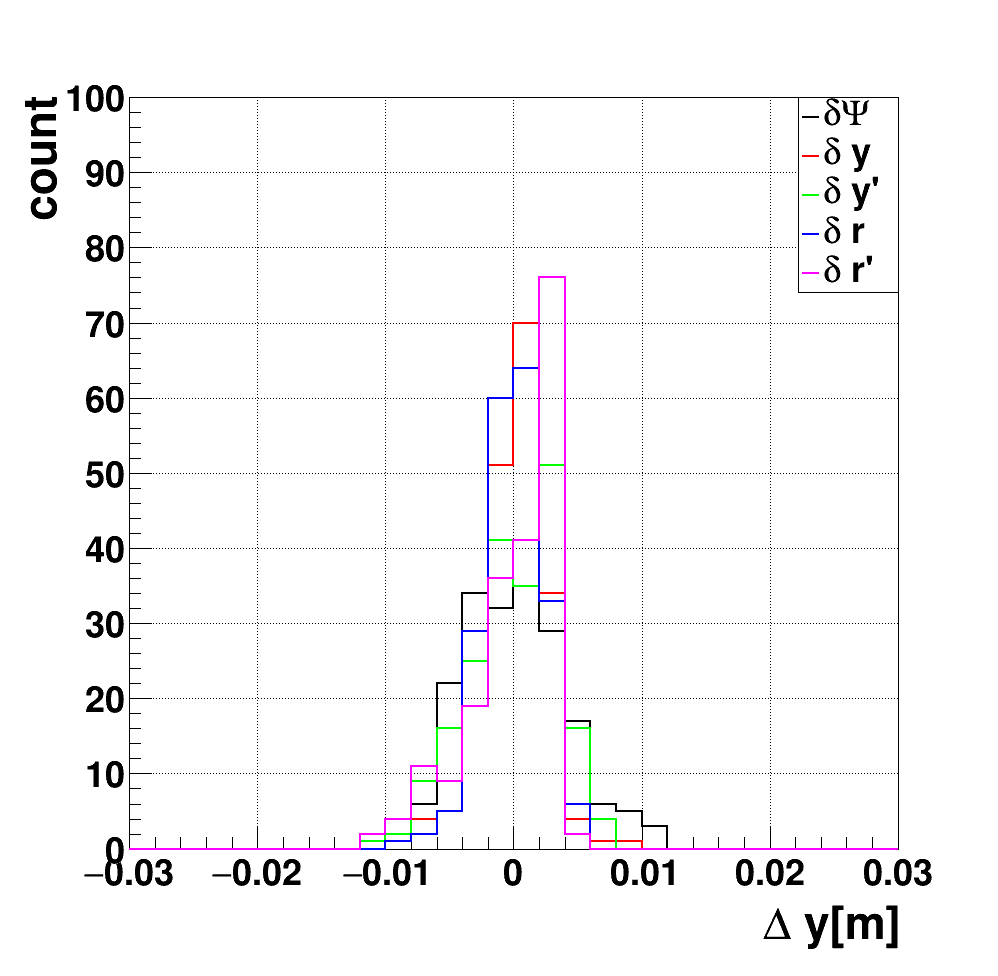}
   \caption{
   One-dimensional histograms of $\Delta y$ obtained by projecting the distributions shown in the lower-left panel of Fig.~\ref{fig:SliceGAt-0} onto the $\Delta y$ axis. The statistical broadening parameters in the $(a_1,b_1)$ space were adjusted for each perturbation family so that the resulting vertical beam distributions at $Y=0$ are confined within approximately $\pm 0.01$~m.}
   \label{fig:disty_aty0}
\end{figure}

Figure~\ref{fig:disty_aty0} shows the one-dimensional histograms of $\Delta y$ obtained by projecting the distributions shown in the lower-left panel of Fig.~\ref{fig:SliceGAt-0} onto the $\Delta y$ axis. The distributions are evaluated at $Y=0$, corresponding to the end point of the injected beam trajectory and the starting point of the backward tracking. 
Although the statistically broadened beams are adjusted to have comparable vertical extents at Y=0, the corresponding spreads in the canonical modal coordinates differ substantially among the perturbation families. In particular, the $\delta y'$- and $\delta r'$-perturbed families require relatively narrow distributions in $b_1$, whereas the $\delta Y$-perturbed family exhibits a somewhat broader distribution. The $\delta R$-perturbed family shows a qualitatively different behavior, with the finite beam extent arising predominantly from the spread along $a_1$, while only a limited spread in $b_1$ is required. Thus, comparable beam sizes in physical space do not correspond to a universal statistical broadening in canonical modal space, but depend on the underlying canonical modal family.

This observation suggests that the respective roles of the $a_k
$- and $b_k$-direction spreads in determining the physical beam extent and the associated eigen-emittances may themselves constitute characteristic properties of each canonical modal family. A systematic investigation of this relation is beyond the scope of the present work and will be addressed in future studies.


Table~\ref{tab:FinitEigenEmittanceMode} represents dominant symplectic eigenmode of $J\Sigma$ from Fig.~\ref{fig:Y_a1_b10}, which is similar to Table~\ref{tab:puturbMode} but for the skeleton beams. 

\begin{table}[!hbt]
   \centering
	\caption{ Symplectic eigenmode of $J\Sigma$ from Fig.~\ref{fig:Y_a1_b10}. This table is compared to Table~\ref{tab:puturbMode}. The second symplectic eigenvectors, together with the three symplectic eigenvalues, are provided in Appendix~\ref{sec:SymplecticEigen} }
      \begin{tabular}{r|c}
       \toprule
	     &Symplectic Eigenmods' vectors  \\
          & [$\Delta~x$,$\Delta~p_x$,$\Delta~y$,$\Delta~p_y$,$\Delta~z$,$\Delta~p_z$]$\times10^{-2}$ \\
       \midrule
    $\delta~y$  $Re(v_1)$ & [ -13.82, 7.55, 31.37, 15.25,  91.88,  9.55 ]\\
      $Im(v_1)$ & [ -1.39, -0.54, -0.60,  0.39,  0.00,  0.02 ] \\ 
 $\|v_1 \|$ &  [  13.89,  7.57,  31.38,  15.26,  91.88,  9.55 ]\\
       \midrule
      $\delta y'$ $Re(v_1)$ &[ -13.50,  7.84,  31.05,  14.90,  92.02,  9.60 ] \\
      $Im(v_1)$&[ -3.29,  0.05, -0.04, -0.24, 0.00,  0.15 ]\\ 
 $\|v_1\|$  &   [  13.89,  7.84,  31.04,  14.90, 92.02,  9.60 ] \\
       \midrule
      $\delta r$~$Re(v_1)$&  [ -55.36,  20.37, -22.82,  6.47, -72.58, -15.40 ]\\
      $Im(v_1)$ &  [ -18.54,  7.64, -6.24,  3.25,  0.00, -1.33] \\
      $\|v\|$  & [  58.40,  21.76,  23.66,  7.24,  72.58,  15.45]\\
      \midrule
 $\delta r'$~$Re(v_1)$&  [  56.40, -2.30,  36.63,  1.80, -73.91,  1.63]\\
      $Im(v_1)$ &  [  1.44, -0.26, -1.05, -0.17, 0.00, -0.38]  \\
      $\|v_1\|$ & [ 56.42,  2.31, 36.64,  1.81,  73.91,  1.67 ]\\
       \bottomrule
   \end{tabular}
   \label{tab:FinitEigenEmittanceMode}
\end{table}

\begin{figure}[!htb]
   \centering  
   \includegraphics*[width=0.7\columnwidth]{figs/Skeleton_ribbon.png}\quad
    \includegraphics*[width=0.7\columnwidth]{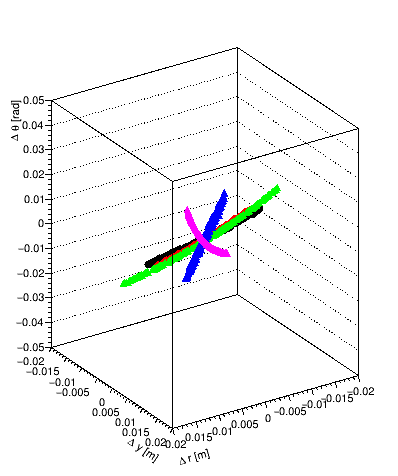}
   \caption{
   Geometrical representation of the statistical beam distributions for the four perturbation families in the reduced ($\Delta~r, ~\Delta~y, ~\Delta~P$) space. 
   The upper panels reproduce the corresponding skeleton-beam representations from Fig.~\ref{fig:r-y-theta_skeleton} for direct comparison. The lower panels show the statistical beam distribution families obtained after Gaussian broadening. Although the beam distributions acquire a finite phase-space extent, the canonical modal topology and the characteristic geometrical differences among the beam families are well preserved.
 The characteristic geometrical differences among the canonical beam families remain clearly identifiable.}
   \label{fig:r-y-theta_skeletonG}
\end{figure}

\begin{figure}[!htb]
   \centering
   \includegraphics*[width=0.7\columnwidth]{figs/Proj_Skeleton_ribbon.png}\quad
   \includegraphics*[width=0.7\columnwidth]{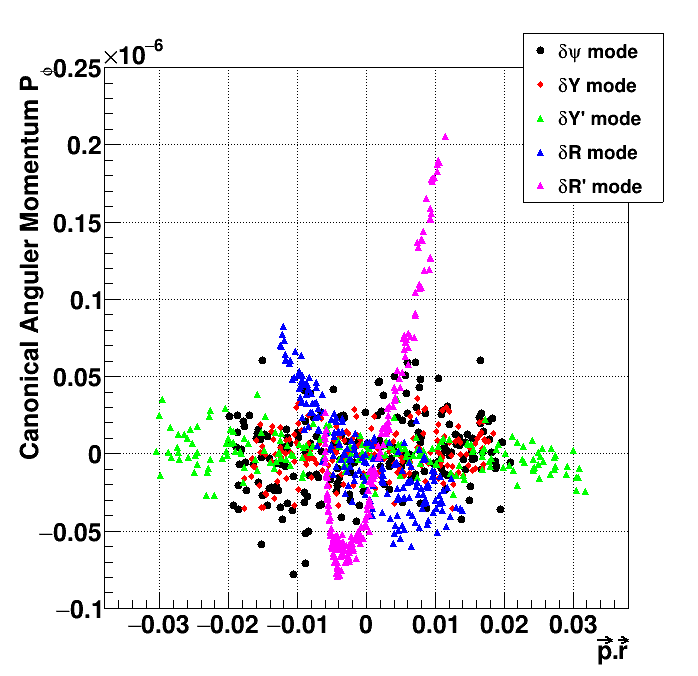}\quad
   \caption{Canonical beam families represented as a function of the inner product between the position and canonical momentum vectors after Gaussian broadening.The upper panels are reproduced from Fig.~\ref{fig:Proj_r-y-theta_skeleton} for direct comparison with the statistical distribution beam distributions shown in the lower panels.
   The lower panels demonstrate that the characteristic separation between the vertical ($\delta~y$,~$\delta~y'$) and radial ($\delta~r$,~$\delta~r'$) perturbation families is preserved after introducing finite emittance.
The scalar representation provides a clear comparison of statistical distribution beam distributions while maintaining the canonical modal classification established in Chapter~\ref{sec:chap3}}
   \label{fig:Proj_r-y-theta_skeletonG}
\end{figure}

The corresponding $r$-$y$-$\vec{r}\cdot\vec{P}$ distributions is shown in Fig.~\ref{fig:r-y-theta_skeletonG}. This is a comparable picture with the canonical skeletons presented previously in Fig.~\ref{fig:r-y-theta_skeleton}. Finally, Fig.~\ref{fig:Proj_r-y-theta_skeletonG} compares the canonical angular momentum as a function of the modal inner product, corresponding to the skeleton results shown in Fig.~\ref{fig:Proj_r-y-theta_skeleton}.

Comparison with the canonical skeletons in Chapter~\ref{sec:chap3} shows that the overall phase-space geometry is largely preserved after statistical broadening. The principal effect of the statistical modal spreads is not to modify the canonical beam topology itself, but to broaden the distribution of canonical angular momentum around the reference trajectories.

\subsection{Canonical Properties of Statistically Broadened Beams}

The statistically broadened beam distributions constructed in the previous section retain the canonical modal structures identified for the corresponding skeleton trajectories. 

Although the introduced statistical spreads broaden the canonical angular-momentum distribution, the characteristic topology of each perturbation family is preserved. In particular, the single-mode and mixed-mode classifications established in Chapter~\ref{sec:chap3} remain clearly distinguishable after particle tracking. These results indicate that the canonical skeleton provides a robust framework for describing realistic beam distributions with finite phase-space extent.

The lower-right panel of Fig.~\ref{fig:SliceGAt-3} shows that the injected statistically broadened beams exhibit little correlation between $\Delta Y$ and the canonical angular-momentum deviation $\Delta P_{\phi}$. After transport, however, the rot-family perturbations ($\delta\psi$, $\delta y$, and $\delta y'$ modes) develop a clear correlation in the $(\Delta Y,\Delta P_{\phi})$ plane, as shown in the corresponding panel of Fig.~\ref{fig:SliceGAt-0}, whereas the radial perturbations retain broader distributions.
This evolution demonstrates that the canonical angular-momentum distribution develops a characteristic structure within the rot-family perturbations during transport, while the overall modal topology is preserved.

The preservation of these canonical properties suggests that statistically broadened beams can be characterized by their canonical modal composition, represented by the complex symplectic eigenvectors, rather than by geometrical beam profiles alone. The role of this canonical modal representation in beam synthesis is examined in the following chapter.

The beam synthesis presented in this chapter is based on statistical spreads introduced around the canonical modal skeletons while preserving their underlying symplectic structure. The resulting beam distributions retain the characteristic phase-space geometry of the corresponding modal families, while the canonical angular-momentum distribution is broadened around the reference trajectories. The present analysis focuses primarily on the symplectic eigenvectors, which define the canonical modal families and their characteristic phase-space geometry. The corresponding symplectic eigenvalues are also evaluated as part of the $J\Sigma$ decomposition; however, no direct physical correspondence between these eigenvalues and the statistical spreads introduced around the canonical skeletons is assumed in the present work. A systematic investigation of this relation is left for future work.

\section{Discussion}\label{sec:chap5}


Conventional beam matching is commonly formulated using Twiss parameters, covariance matrices, and related quantities such as BMag~\cite{BMAG}. These methods quantify the agreement of beam envelopes or projected phase-space ellipses and provide effective criteria for optical matching in beam transport systems.

The present framework addresses a different design problem. In three-dimensional spiral injection, the objective is not only to reproduce projected beam sizes but also to preserve the canonical modal structure associated with coherent beam rotation. 
Consequently, the similarity measures $C_6$ introduced in Chapter~\ref{sec:chap3} characterize canonical modal matching rather than conventional optical matching.

From this viewpoint, Twiss parameters, BMag, and $C_6$ should not be regarded as competing quantities but as indicators describing different aspects of beam design. Twiss parameters and BMag quantify optical matching of projected beam distributions, 
whereas $C_6$ characterizes the similarity between canonical modal skeletons through the overlap of their dominant symplectic eigenvectors. It therefore describes modal orientation, but does not by itself characterize the phase-space distribution of particles around the modal skeleton. In contrast, the projection residual $\eta$ quantifies how individual phase-space points are represented by a prescribed canonical modal plane. For a broadened beam distribution, the relevant information is therefore not only the magnitude of $\eta$, but also its distribution in phase space. Such an $\eta$-distribution provides a complementary measure of the compatibility between an actual beam distribution and a canonical modal family. An application of this representation to a model-case beam distribution is presented in Appendix~\ref{sec:E34}.

\section{Summary and outlook}
In this work, three-dimensional spiral injection has been reformulated in six-dimensional canonical phase space using the eigensystem of $J\Sigma$.

The analysis further shows that the canonical beam families can be naturally classified according to the response of the canonical angular momentum. The rotationally symmetric beam and perturbations along the solenoid axis ($\delta~y$ and $\delta~y'$) preserve the canonical angular momentum to first order and are therefore represented by single-mode canonical families. 
In contrast, radial perturbations ($\delta~r$ and $\delta~r'$) exhibit strong correlations between the canonical angular momentum and the spatial dependence of the magnetic field, resulting in dynamically displaced mixed-mode families. 
These observations reveal two distinct classes of canonical beam behavior in three-dimensional spiral injection.

The canonical modal skeletons were further extended to statistically broadened beam distributions by introducing statistical spreads around the modal coordinates. The resulting distributions retain the characteristic phase-space geometry and modal classification of the corresponding skeleton families after particle tracking. This demonstrates that the canonical skeletons provide a practical basis for constructing beam distributions with finite phase-space extent while preserving their underlying modal structure.

To visualize these canonical structures, the reduced ($\Delta~r$,~$\Delta~y$,~$\Delta~(\vec{r}\cdot\vec{P})$) representation was introduced. Although all beam families appear as remarkably flattened distributions compared with conventional beam phase-space distributions, this geometry is shown to be a direct consequence of the angular-momentum constraint imposed by an approximately axisymmetric solenoidal magnetic field. The ($\Delta~r$,~$\Delta~y$,~$\Delta~(\vec{r}\cdot\vec{P})$) representation therefore provides an intuitive visualization of the canonical beam structure while remaining consistent with the underlying six-dimensional modal description.

An important remaining question is whether practical injection beams can be designed to satisfy the common canonical properties shared by these three scenarios while maintaining high injection efficiency in realistic magnetic fields. This question will be addressed in a subsequent paper through numerical optimization and experimental validation using realistic magnet-field distributions.
An application of the present framework to an optimized injection-beam distribution is presented in Appendix~\ref{sec:E34}.

\textbf{Outlook.}
The present framework is not limited to the beam distributions investigated in this work. Because beam families are represented by canonical modal coordinates, the same formulation can be extended to systematic beam
synthesis, optimization, and AI-assisted beam design for
general coupled beam transport systems.

\begin{acknowledgments}
The author would like to thank Dr. Oide and Dr. Furukawa
for their valuable comments encouraging a more intuitive
reformulation of the physical interpretation of
three-dimensional spiral injection, which motivated the
development of the present work.

The author also thanks Dr. Matsushita for many helpful
discussions throughout this study.
\end{acknowledgments}

\clearpage
\appendix
\section{Canonical Coordinates in Realistic Solenoidal Fields}\label{sec:Aphi}

\begin{figure}[!htb]
   \centering
   \includegraphics*[width=.9\columnwidth]{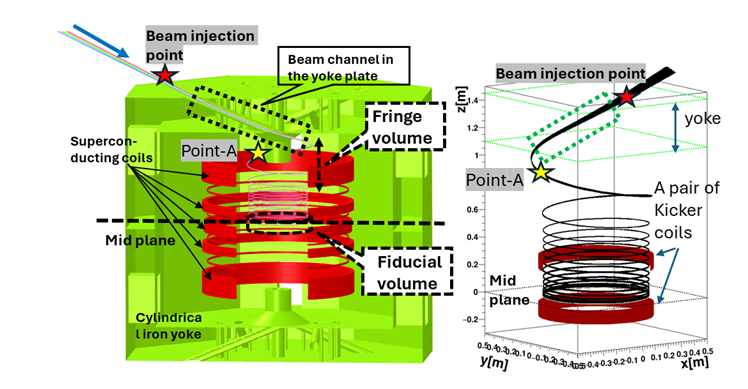}
   \caption{
   Overview of the three-dimensional spiral injection beamline used in the J-PARC E34 demonstration experiment. The figure illustrates the overall injection system, including the injection channel, pulsed kicker magnets, and compact solenoidal storage region. The beamline configuration serves as the basis for the canonical analysis presented in this Appendix. Adapted from~\cite{Iinuma:ipac2026-thp5621}}
   \label{fig:THP5621_fig1}
\end{figure}

The canonical coordinates used throughout this work are constructed from an axisymmetric representation of the three-dimensional magnetic field calculated with OPERA.
The reconstruction procedure follows the methodology developed for the storage magnet design shown in Fig.~\ref{fig:THP5621_fig1} in the next section~\ref{sec:E34}~\cite{Abe:NIMA890}.

\subsection{Axisymmetric Magnetic-Field Representation}

\begin{figure}[!htb]
   \centering
   \includegraphics*[width=0.45\columnwidth]{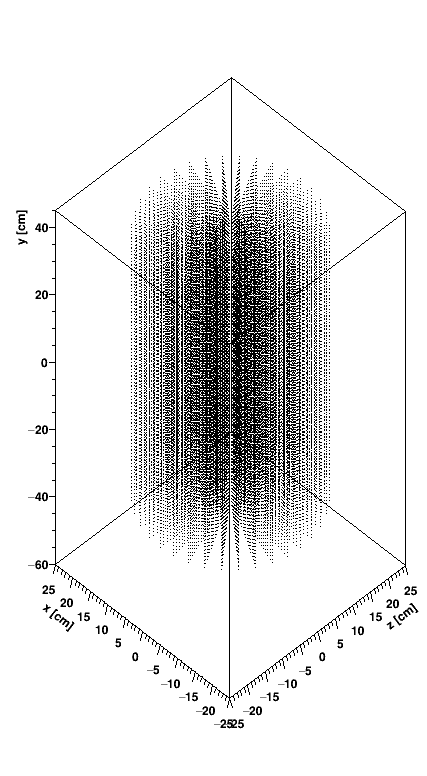}\quad
   \includegraphics*[width=0.48\columnwidth]{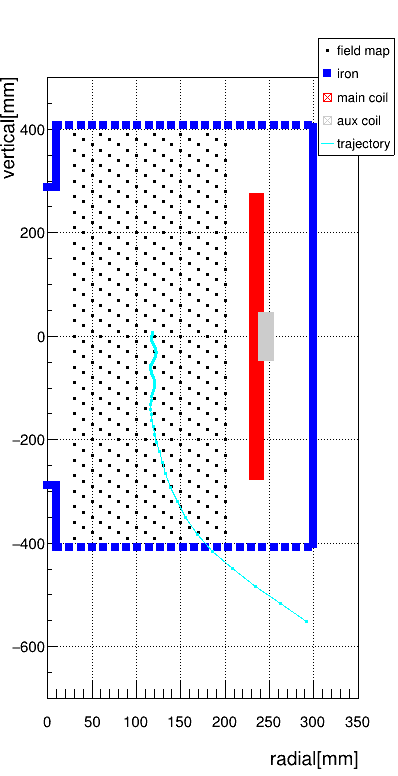}\quad
   \caption{
   Construction of an axisymmetric magnetic-field representation from the three-dimensional OPERA field model. Left: magnetic-field sampling on 36 azimuthal planes at 10° intervals. Right: meridional cross section of the storage magnet showing the sampled field region, iron yoke, main and auxiliary coils, and a representative beam trajectory. The sampled fields are azimuthally averaged and subsequently reconstructed by an equivalent circular-current model to evaluate the vector potential $A_{\phi}$.}
   \label{fig:SITE_MAP}
\end{figure}

To obtain an axisymmetric field suitable for canonical-coordinate calculations, the three-dimensional magnetic field is sampled on a set of azimuthal planes at intervals of $10^\circ$, as illustrated in left panel of Fig.~\ref{fig:SITE_MAP}. The sampling points cover the storage region inside the iron yoke, where the magnetic field is designed to satisfy approximate axial symmetry. The magnetic-field components are averaged over the 36 azimuthal planes to construct a two-dimensional magnetic-field map in the (r,Y) plane, thereby suppressing small numerical asymmetries originating from the finite-element discretization of the OPERA model.

Following the procedure described in Ref.~\cite{Abe:NIMA890}, the averaged magnetic field is approximated by an equivalent circular-current model. Filament-loop currents are distributed according to the geometry of the iron yoke, the main solenoid, and the auxiliary coils. Their amplitudes are determined using truncated singular-value decomposition (tSVD), yielding a smooth axisymmetric magnetic-field representation. From this equivalent current distribution, the azimuthal vector potential $A_\phi(r,Y)$ is evaluated and subsequently used to construct the canonical momentum introduced in the following section.

\subsection{Canonical Momentum}
The axisymmetric magnetic-field representation described in the previous section provides the azimuthal vector potential $A_{\phi}$, from which the canonical momentum is constructed.

For an axisymmetric magnetic field, the vector potential is expressed as
\begin{equation}
    A=(0,A_\phi, 0)
\end{equation}
where the radial and axial components vanish under the adopted gauge. The canonical momentum is therefore defined by
\begin{equation}
    P=p+qA
\end{equation}
where $p$ denotes the mechanical momentum and $q$ is the particle charge.

In cylindrical coordinates, the canonical angular momentum is written as
\begin{equation}
    P_{\phi}=m\gamma r^2\dot{\phi}+qA_{\phi}(r,Y),
\end{equation}
where the first term represents the mechanical angular momentum and the second term is the contribution from the vector potential.

Throughout this work, the canonical angular momentum is evaluated from the reconstructed axisymmetric vector potential and used as one of the canonical phase-space variables for constructing the covariance matrix $\Sigma$. The canonical coordinates obtained in this manner are subsequently employed for the eigenmode analysis of $J\Sigma$ presented in the main text and Appendix~\ref{sec:E34}.

\subsection{Numerical Implementation}

Particle trajectories are calculated using the three-dimensional OPERA magnetic-field map. The trajectory integration is performed in Cartesian coordinates using a sixth-order Runge--Kutta method. When required for the canonical analysis, the particle coordinates and momenta are transformed between Cartesian and cylindrical coordinate systems by the corresponding coordinate transformation.

Using the axisymmetric vector potential $A_{\phi}(r,Y)$ obtained in Section A.1, the canonical momentum is evaluated for each particle. The resulting six-dimensional canonical phase-space coordinates are then used to construct the covariance matrix $\Sigma$, from which the matrix $J\Sigma$ is formed for the canonical modal analysis presented in the main text and Appendix~B.

\begin{figure}[!htb]
   \centering
      \includegraphics*[width=0.48\columnwidth]{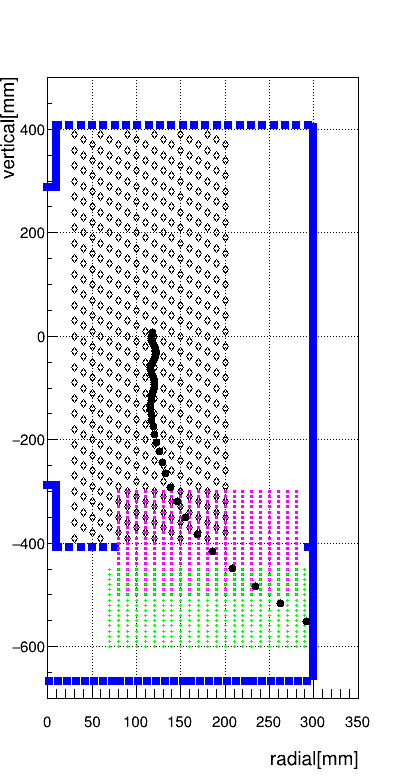}\quad
   \caption{
   Field-map region and representative injection trajectory extending from the axisymmetric region enclosed by the iron yoke toward the external region. The present analysis evaluates the canonical momentum within the axisymmetric region. Extension of the vector-potential and canonical-momentum evaluation along the incoming trajectory beyond this region is left for future work.
   }
   \label{fig:SITE_MAP_PASJ}
\end{figure}

The present analysis is restricted to the axisymmetric region enclosed by the iron yoke. Extension of the canonical-momentum evaluation to the non-axisymmetric region outside the magnet as in Fig.~\ref{fig:SITE_MAP_PASJ}, relevant to the incoming injection trajectory, is left for future work.

\clearpage
\section{Relation to PCA and SVD}\label{sec:SVD}
The SVD considered in this appendix refers to the decomposition of the centered particle-data matrix and is distinct from the tSVD used for the magnetic-field reconstruction in Appendix~\ref{sec:Aphi}.

The relation between the symplectic modal analysis based on ($J\Sigma$) and conventional principal-component analysis (PCA) or singular-value decomposition (SVD) can be stated explicitly. Let ($X$) denote a centered particle-data matrix, whose rows contain the (2n)-dimensional canonical phase-space coordinates of the particles. The covariance matrix is

\begin{equation}
\Sigma=\frac{1}{N-1}X^{T}X .
\end{equation}

For the singular-value decomposition
\begin{equation}
    X=USV^{T},
\end{equation}
the covariance matrix can be written as
\begin{equation}
\Sigma
=V\frac{S^{2}}{N-1}V^{T}.
\end{equation}

Thus, the right singular vectors of the centered particle-data matrix are the eigenvectors of ($\Sigma$), i.e., the PCA directions, apart from sign conventions and rotations within degenerate subspaces. A truncated SVD therefore corresponds to retaining a subset of the variance-ranked PCA directions. In a six-dimensional phase space, the complete PCA/SVD representation contains up to six real directions.

The symplectic modal analysis considered in this work decomposes a different matrix structure. With the canonical symplectic matrix

\begin{equation}
 J=
   \begin{pmatrix}
     J_2&0&0\\
     0&J_2&0\\
     0&0&J_2
   \end{pmatrix},
 \qquad
  J_2=
  \begin{pmatrix}
    0&1\\
    -1&0
   \end{pmatrix},
\end{equation}

the modal structure is obtained from

\begin{equation}
    J\Sigma v_k=\lambda_k v_k.
\end{equation}

For a positive-definite covariance matrix in a six-dimensional canonical phase space, the eigenvalues occur in three conjugate pairs,

\begin{equation}
    \lambda_k=\pm i\epsilon_k,
\qquad k=1,2,3 ,
\end{equation}

where ($\epsilon_k$) characterize the three symplectic modes. Each independent complex eigenvector may be written as

\begin{equation}
    v_k=\mathrm{Re}(v_k)+i\mathrm{Im}(v_k),
\end{equation}

and the complete set of three modes provides a six-dimensional real representation through their real and imaginary components.

Accordingly, retaining all three symplectic modes is dimensionally comparable to retaining all six real PCA/SVD directions. This correspondence in dimensionality should not, however, be interpreted as an equivalence of the two decompositions. PCA/SVD diagonalizes the covariance structure in a Euclidean metric and orders orthogonal directions according to statistical variance. The ($J\Sigma$) eigenproblem instead incorporates the symplectic matrix ($J$) and organizes the distribution into modes associated with the canonical phase-space structure.

In particular, an individual symplectic eigenvector cannot in general be identified with an individual PCA or SVD vector. Relations between a PCA direction and a quantity derived from a symplectic eigenvector, such as ($J\mathrm{Re}(v_k)$), are therefore not assumed to constitute a general mathematical identity. When such a correspondence occurs, it must be established for the particular covariance structure under consideration.

The two representations therefore describe different structures of the same statistical distribution.
PCA/SVD identifies variance-ranked Euclidean directions, whereas the ($J\Sigma$) analysis identifies the modal organization associated with the symplectic structure of canonical phase space. The application presented in Appendix~\ref{sec:E34} provides an explicit test of these relations for the particle distributions considered in this work.

\clearpage

\section{Case Study: Canonical Interpretation of a J-PARC Muon $g$-2/EDM Injection Beam}\label{sec:E34}
As a case study, the injection beam model developed for the J-PARC muon g-2/EDM experiment is analyzed. The experiment employs a 3 T superconducting storage solenoid based on MRI technology, providing a highly axisymmetric magnetic field with a local field uniformity better than 0.1 ppm to enable precision measurements of the muon spin precession. The injection optics were designed to suppress the vertical betatron oscillation (VBO) in the storage region, and successful injection is defined by a maximum VBO amplitude smaller than 50 mm, following the criterion adopted in the original beam-design study. 
Owing to the axisymmetric nature of the storage field, the canonical framework developed in the main text is directly applicable to analyze the optimized injection beam.

To maintain consistency with the notation used throughout this paper, the solenoid axis is denoted by the Y axis in this Appendix, although the original coordinate of this system used the Z axis.


\subsection{Canonical Modal Families Generated by Reverse Tracking}\label{sec:E34skelton}
The reverse-tracking procedure follows the same framework introduced in the main text. Reverse tracking starts from the storage reference orbit at Y=0, where the stored muon beam performs precision spin-precession measurements in the 3 T superconducting storage magnet. The reverse trajectories are propagated through the storage region including the pulsed kicker field.

The canonical modal analysis presented in this Appendix is performed at Y=0.95 m. This position is located inside the iron yoke of the storage magnet, where the magnetic field remains sufficiently axisymmetric. Beyond this point, the influence of the yoke structure becomes significant, and the magnetic field gradually deviates from axial symmetry. The present analysis is therefore restricted to the axisymmetric region, where the canonical formalism developed in the main text is directly applicable.

\begin{figure}[!htb]
   \centering
   \includegraphics*[width=.9\columnwidth]{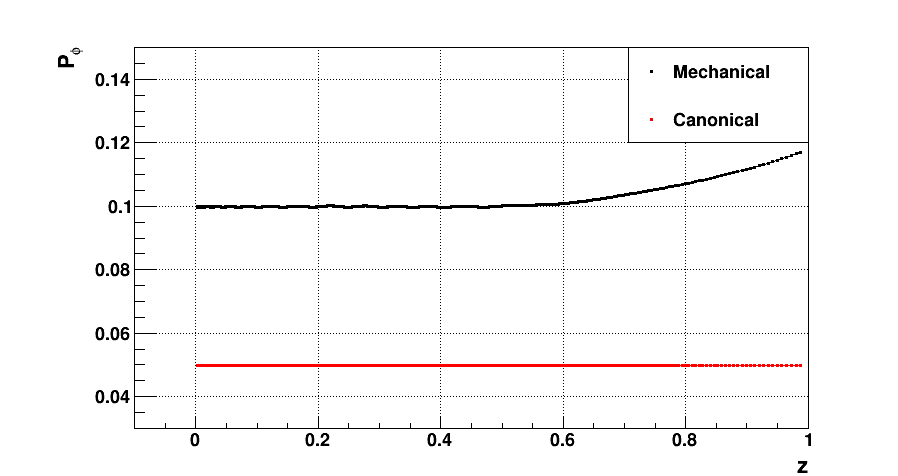}
   \caption{Mechanical and canonical angular momenta of the reverse-tracked reference particle as a function of the longitudinal coordinate. Although the mechanical angular momentum changes after the kicker owing to the electromagnetic interaction, the canonical angular momentum remains conserved within numerical precision throughout the axisymmetric region. This conservation property justifies the application of the canonical modal analysis presented in this Appendix.
   }
   \label{fig:E34Phi0Phi}
\end{figure}

Figure~\ref{fig:E34Phi0Phi} compares the mechanical and canonical angular momenta along the reverse-tracked reference trajectory. Although the mechanical angular momentum changes after the kicker, reflecting the electromagnetic interaction, the canonical angular momentum remains conserved within numerical precision throughout the axisymmetric region. This conservation property provides the physical basis for applying the canonical modal analysis to the reverse-tracked beam.

The reverse tracking is subsequently continued from the analysis point to the external injection point outside the storage magnet. The upstream trajectory calculation and the injection optimization have already been described in previous publications. In this Appendix, only the canonical interpretation of the optimized injection beam is discussed.

Following the procedure introduced in the main text, four perturbative skeletons, denoted by Y, Y', R and R' are generated by applying small perturbations at the end point of beam injection (Y=0), immediately after the kicker. Twenty reverse-tracked trajectories are calculated for each perturbation family.

\begin{figure}[!htb]
   \centering
   \includegraphics*[width=.9\columnwidth]{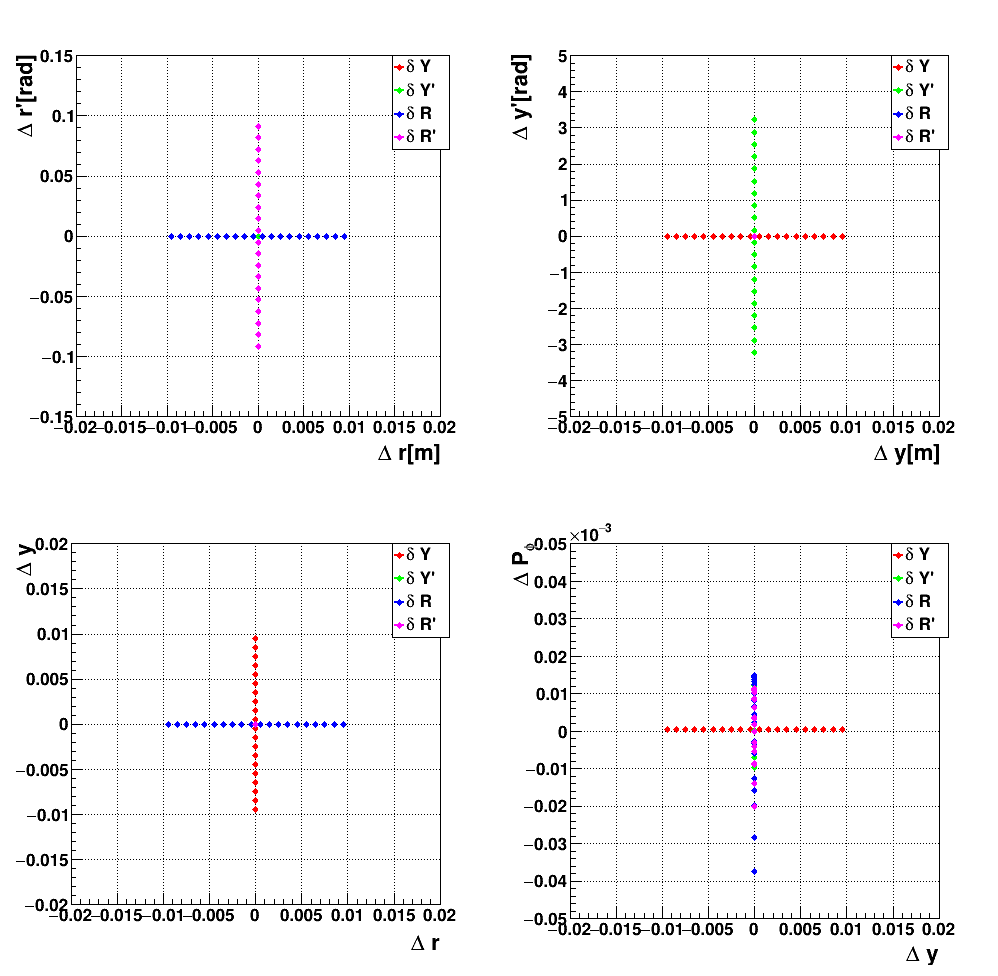}
   \caption{
   Four perturbation families generated by backward tracking from the end point of beam injection (immediately after the kicker). 
   The perturbations correspond to vertical position (Y), vertical momentum (Y'), radial position (R), and radial momentum (R'). 
   The lower-right panel shows the corresponding canonical angular momentum of each perturbation family. 
   As in the canonical examples presented in the main text, the Y and Y' perturbations preserve nearly constant canonical angular momentum, whereas the R and R'~perturbations generate systematic variations.}
   \label{fig:E34_perturb0}
\end{figure}

\begin{figure}[!htb]
   \centering
   \includegraphics*[width=.9\columnwidth]{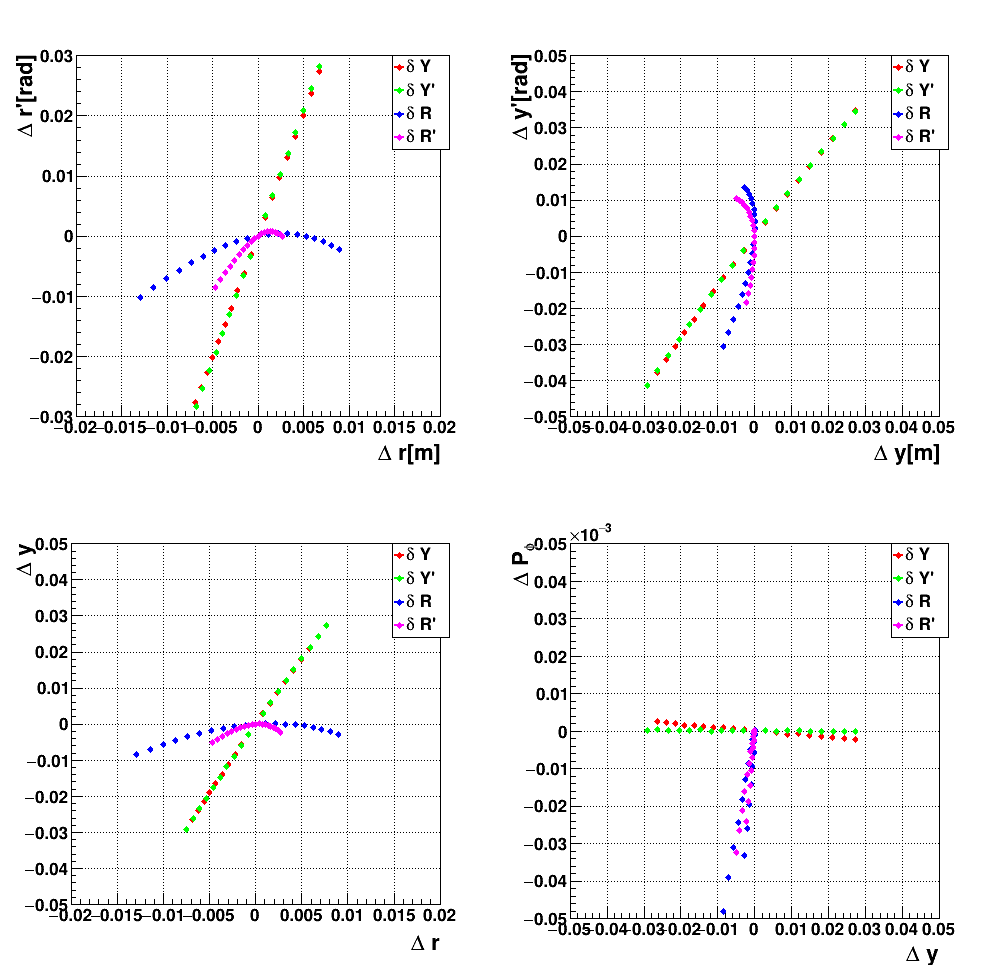}
   \caption{
   Beam phase-space distributions of the four perturbation families at the analysis point (Y=0.95 m). The four panels show the transverse phase-space projections (r,r'), (y,y'), (r,y), and (y,$P_{\phi}$). Beam slices are extracted on the plane normal to the momentum vector of the reference trajectory, following the same procedure as described in the main text. The resulting beam distributions are used for covariance-matrix construction and canonical modal analysis.}
   \label{fig:E34_perturb95}
\end{figure}

Figure~\ref{fig:E34_perturb0} illustrates the initial perturbations defined at Y=0, while Fig.~\ref{fig:E34_perturb95} shows the corresponding beam slices after transport to the analysis point (Y=0.95 m).
Owing to the nonlinear magnetic transport through the storage field and the kicker, the initially orthogonal perturbations evolve into distinct canonical modal families.

\begin{figure}[!htb]
   \centering
   \includegraphics*[width=.7\columnwidth]{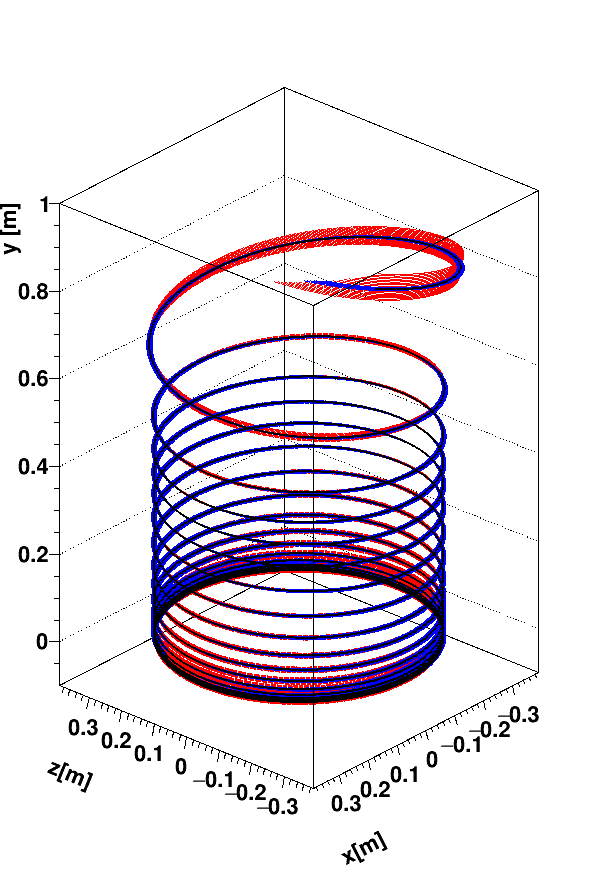}
   \caption{
   Representative three-dimensional backward-tracked trajectories for the Y~(red)~and R~(blue) perturbation families. Beam slices are extracted on the plane normal to the momentum vector of the reference trajectory at Y=0.95 m, following the same procedure as described in the main text.}
   \label{fig:E34-3D}
\end{figure}

As shown in the main text, perturbations in the R and R' directions generate a distribution of canonical angular momentum, whereas the Y and Y' perturbations preserve nearly the same canonical angular momentum. Figure~\ref{fig:E34-3D} shows the corresponding three-dimensional backward-tracked trajectories for the Y- and R-perturbed beams.

\begin{figure}[!htb]
   \centering
   \includegraphics*[width=.9\columnwidth]{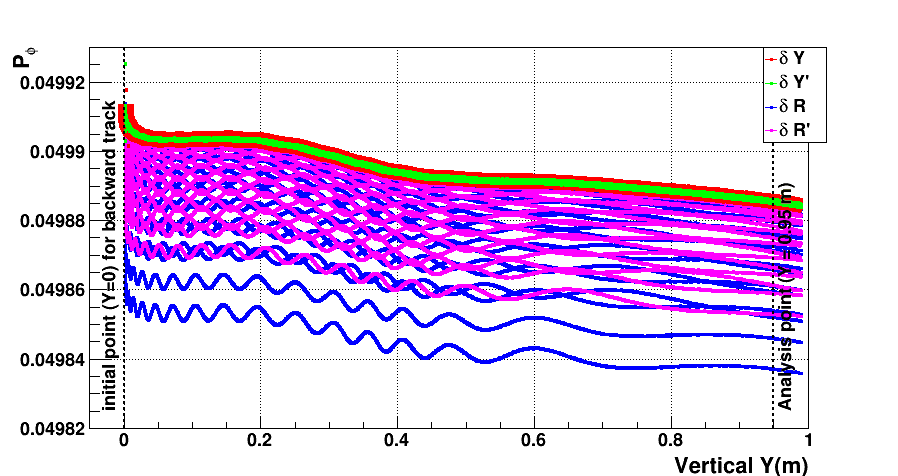}
   \caption{
   Evolution of the canonical angular momentum $P_{\phi}$ for the four perturbation families during backward tracking. The trajectories start from the end point of beam injection (Y=0) and terminate at the analysis point (Y=0.95 m). The Y and Y' perturbations preserve nearly constant canonical angular momentum, whereas the R and R' perturbations show systematic variations.}
   \label{fig:E34Phi}
\end{figure}

Figure~\ref{fig:E34Phi} presents the evolution of the canonical angular momentum of all reverse-tracked trajectories as a function of the solenoid-axis coordinate. For each individual trajectory, the canonical angular momentum remains conserved within approximately $10^{-5}$ throughout the axisymmetric transport region. 
The Y and Y' perturbation families share nearly the same canonical-angular-momentum values, whereas the R and R' perturbations generate systematic variations among the trajectories.

At the analysis point (Y = 0.95 m), beam slices are extracted from each perturbation family. 
 Figure~\ref{fig:E34_perturb95} summarizes the corresponding phase-space distributions.
These beam slices constitute the covariance matrices analyzed in the following section. The matrices $J\Sigma$ are constructed from these phase-space distributions, and their eigenmodes are subsequently evaluated using the canonical modal analysis developed in the main text.

Consistent with the canonical examples presented in the main text, the Y and Y' perturbation families exhibit nearly identical beam phase-space distributions, whereas the R and R' perturbation families constitute a separate canonical family with distinct phase-space characteristics.

Table~\ref{tab:E34SkeletonigenEmiRatio} summarizes the relative magnitudes of the symplectic eigenvalues of $J\Sigma$, showing the dominance of the first mode.
\begin{table}[!hbt]
   \centering
	\caption{ Ratio of Symplectic eigenmode values of $J\Sigma$ from Fig.~\ref{fig:E34_perturb95}. }
   \begin{tabular}{l|cl}
       \toprule
	     &Ratio & $\epsilon_2$/$\epsilon_1$ \\
       \midrule
      $\delta y$&$ 10^{-4}$&1.12$\times10^{-9}$/1.40$\times10^{-5}$ \\
      $\delta y'$& $10^{-4}$&1.26$\times10^{-9}$/1.96$\times10^{-5}$\\
      $\delta r$& $10^{-3}$&8.68$\times10^{-9}$/2.84$\times10^{-6}$ \\
      $\delta r'$&$10^{-2}$&7.66$\times10^{-8}$/1.67$\times10^{-6}$ \\
       \bottomrule
   \end{tabular}
   \label{tab:E34SkeletonigenEmiRatio}
\end{table}

 Table~\ref{tab:E34SkeletonigenEmittanceMode} summarizes the symplectic eigenvalues and eigenvectors of $J\Sigma$, while Table~\ref{tab:E34SkeletonPCA} lists the corresponding principal components obtained by PCA of $\Sigma$. 
 A close correspondence between the PCA directions and $J\mathrm{Re}(v_k)$, analogous to that found in the canonical examples of the main text, is also observed for these perturbative skeletons.

\begin{table}[!hbt]
   \centering
	\caption{ Symplectic eigenmode of $J\Sigma$ from Fig.~\ref{fig:E34_perturb95}. }
   \begin{tabular}{l|lc}
       \toprule
	     id$\#$&&Symplectic Eigenmods' vectors \\
           &&[$\Delta~x$,$\Delta~p_x$,$\Delta~y$,$\Delta~p_y$,$\Delta~z$,$\Delta~p_z$]$\times10^{-2}$ \\
       \midrule
      $\delta~y$&$Re(v_1)$ &[83.09,23.67,-21.38,18.49,-24.17,-33.77] \\
      &$Im(v_1)$ &[0.00,0.40,-2.56,-0.48,-1.96,-1.14]   \\ 
       &$Re(v_2)$ &[-44.47,-16.67,30.14,12.98,61.20,37.22]   \\
        &$Im(v_2)$ &[-37.20,-8.01,-0.41,-0.20,0.00,9.38]   \\
         \midrule
      $\delta y'$&$Re(v_1)$ &[83.21,23.86,-22.23,18.93,-23.59,-33.00] \\
      &$Im(v_1)$ &[0.00,0.55,-2.82,-0.50,-1.97,-1.13] \\ 
     &$Re(v_2)$ &[64.21,-18.93,20.34,10.60,44.13,33.36] \\
      &$Im(v_2)$&[0.00,4.42,-17.60,-8.13,-36.65,-13.90]\\
            \midrule
      $\delta r$&$Re(v_1)$&[67.30,41.92,45.02,4.28,-36.84,-13.89] \\
      &$Im(v_1)$&[0.00,-2.25,-10.17,1.64,1.65,-2.23]   \\
            \midrule
     $\delta r'$&$Re(v_1)$ &[41.86,-28.08,-16.23,-72.02,38.94,-5.69] \\
      &$Im(v_1)$&[19.44,3.60,-4.26,0.00,-2.38,-6.73]   \\ 
       \bottomrule
   \end{tabular}
   \label{tab:E34SkeletonigenEmittanceMode}
\end{table}

~
\begin{table}[!hbt]
   \centering
	\caption{ PCA of $\Sigma$ from Fig.~\ref{fig:E34_perturb95}.}
   \begin{tabular}{l|lc}
       \toprule
	     id$\#$&&PCA values \\
         &&[$\Delta~x$,$\Delta~p_x$,$\Delta~y$,$\Delta~p_y$,$\Delta~z$,$\Delta~p_z$]$\times10^{-2}$ \\
       \midrule

     $\delta~y$&$u_1$ & [23.67,-83.16,18.49,21.32,-33.79,24.20]\\
      &$u_2$ &[-4.38,-37.24,24.91,-69.10,21.01,-52.51]   \\ 
         \midrule
      $\delta y'$&$u_1$ &[23.79,-83.29,18.92,22.17,-33.01,23.62] \\
      &$u_2$ &[-5.57,-36.76,24.65,-72.16,19.04,-49.41] \\ 
       \midrule
      $\delta r$&$u_1$&[42.00,-65.02,3.06,-50.04,-12.035,36.73] \\
      &$u_2$& [5.03,53.65,-22.32,-74.10,33.26,1.03]  \\
            \midrule
     $\delta r'$&$u_1$ &[25.51,51.39,70.35,-18.06,9.18,36.71] \\
      &$u_2$&[31.50,-74.49,35.08,17.07,-30.94,31.29]   \\ 
       \bottomrule
   \end{tabular}
   \label{tab:E34SkeletonPCA}
\end{table}

\subsection{Canonical Analysis of the Optimized Injection Beam}
\begin{figure}[!htb]
   \centering
   \includegraphics*[width=.9\columnwidth]{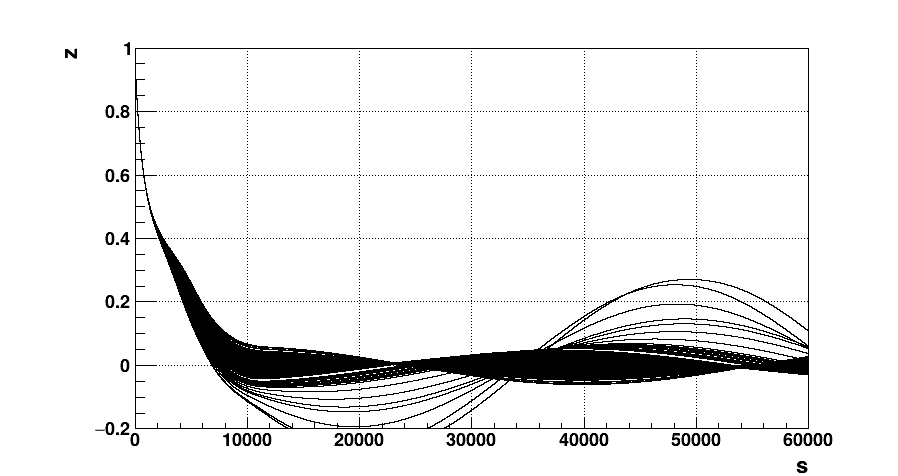}
   \caption{Vertical motion of the optimized injection beam reproduced from the original IPAC beam-design study. The trajectories are obtained by backward tracking from the analysis point (Y=0.95 m), followed by sixth-order Runge--Kutta tracking to the external injection point. After passing through the kicker, the beam performs vertical betatron oscillations in the storage region. The optimized beam satisfies the design criterion of a maximum vertical betatron oscillation (VBO) amplitude smaller than 50 mm.
   }
   \label{fig:THP5621_fig2}
\end{figure}

Before applying the canonical modal analysis, the optimized injection beam is briefly reviewed. Figure~\ref{fig:THP5621_fig2} reproduces the vertical motion of the optimized injection beam obtained in the original beam-design study. The trajectories are calculated by backward tracking from the analysis point through the kicker and subsequently propagated upstream to the external injection point using a sixth-order Runge--Kutta integration. After injection, the beam undergoes vertical betatron oscillations in the storage region, while satisfying the design criterion of a maximum VBO amplitude below 50 mm. In the following analysis, the beam phase space at the analysis point (Y=0.95 m) is interpreted within the canonical modal framework developed in the main text.


\begin{figure}[!htb]
   \centering
   \includegraphics*[width=.9\columnwidth]{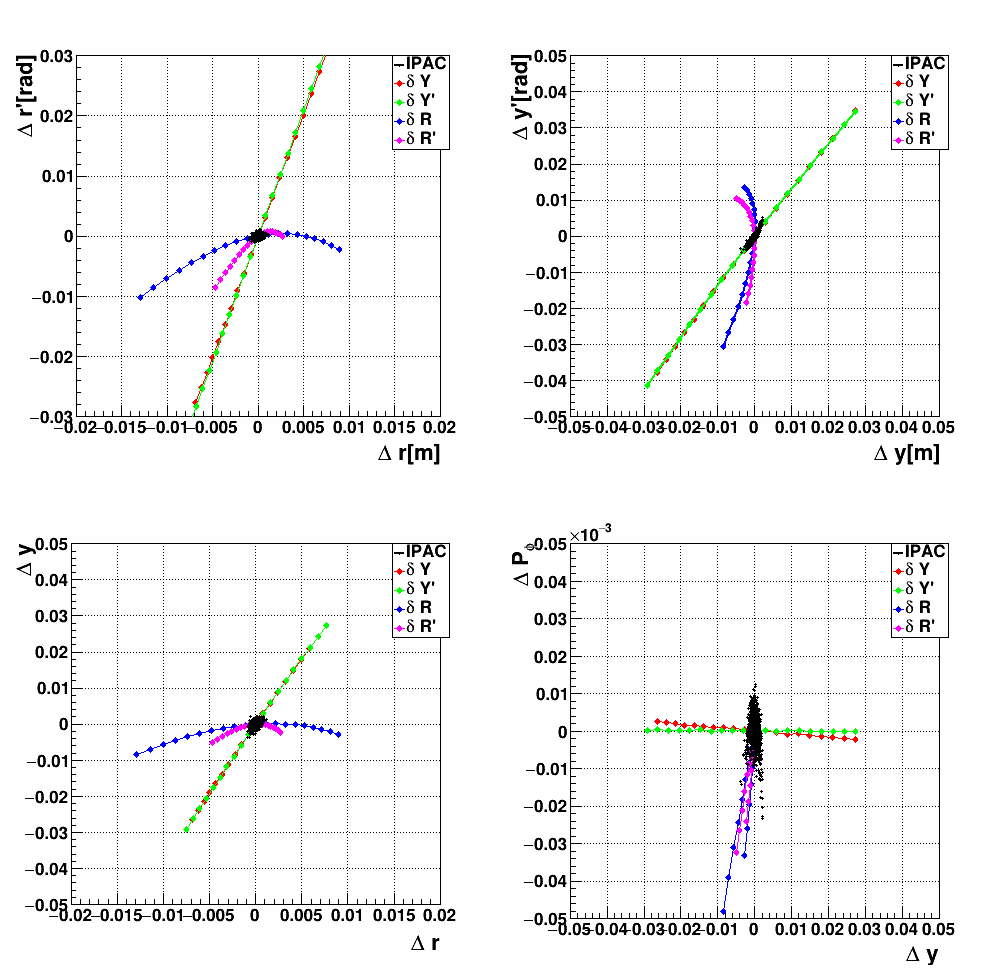}
   \caption{
   Phase-space distributions at the analysis point (Y=0.95 m). The black points denote the optimized injection beam obtained in the original IPAC beam-design study, while the colored curves represent the four perturbative canonical skeletons generated by reverse tracking. The optimized beam occupies only a localized region of the perturbative phase space.The figure is obtained by superimposing the optimized beam distribution (black points) onto the perturbative skeletons shown in Fig.~\ref{fig:E34_perturb95}.}
   \label{fig:E34IPAC26_phase}
\end{figure}

Figure~\ref{fig:E34IPAC26_phase} compares the optimized injection beam at the analysis point (Y=0.95 m) with the perturbative skeletons introduced in Fig.~\ref{fig:E34_perturb95} (Sec.~\ref{sec:E34skelton}). The black points represent the optimized injection beam from the original IPAC study, superimposed on the corresponding perturbative skeletons.

\begin{figure}[!htb]
   \centering
   \includegraphics*[width=.9\columnwidth]{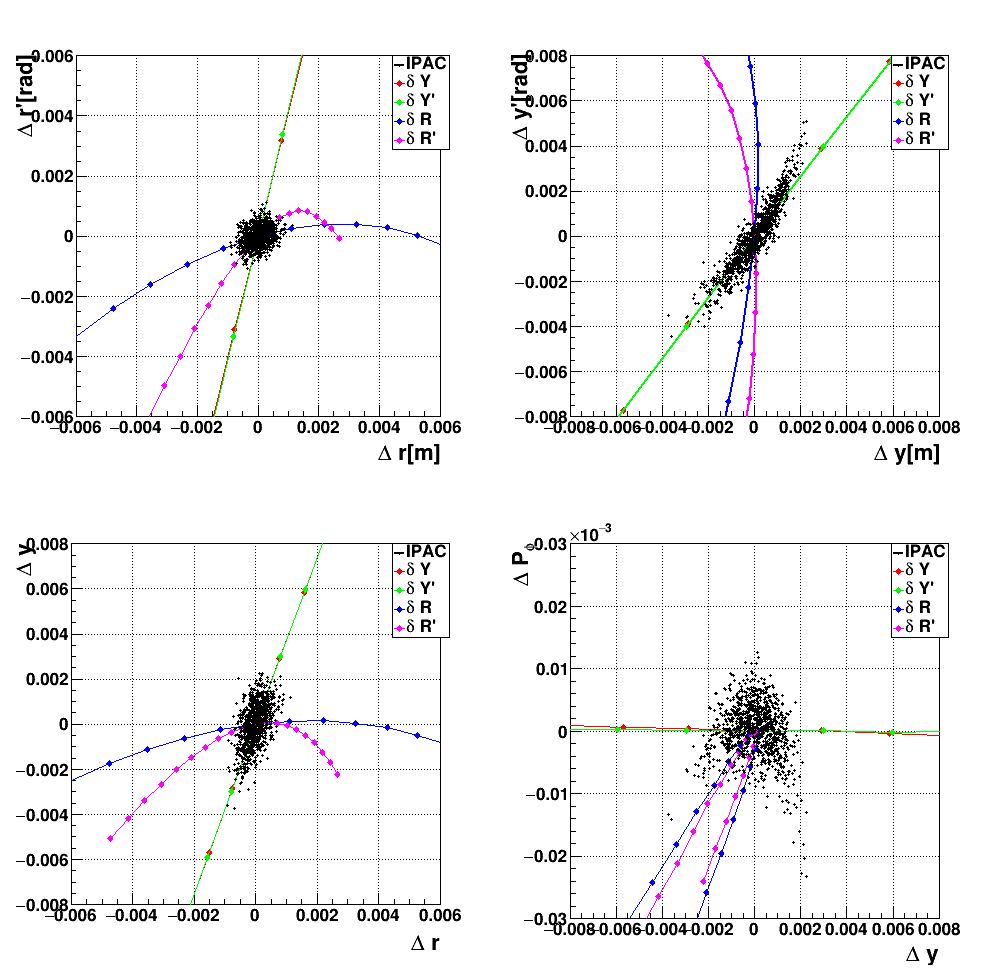}
   \caption{
   Enlarged views of Fig.~\ref{fig:E34IPAC26_phase} around the optimized injection-beam distribution. The perturbative skeletons provide the reference canonical structures for its interpretation in the following canonical modal analysis.}
   \label{fig:E34IPAC26_zoom}
\end{figure}

Figure~\ref{fig:E34IPAC26_zoom} presents enlarged views around the beam core. Although the optimized beam occupies only a limited region of the perturbative phase space, the distributions exhibit clear similarities to the corresponding canonical skeletons. 

Equation~\ref{eq:E34_eps} summarizes symplectic eigenvalues of $J\Sigma$ from a distribution shown in Fig.~\ref{fig:E34IPAC26_zoom}.

\begin{eqnarray}
    \epsilon_1&=& 2.44\times 10^{-7}  \nonumber \\
   \epsilon_2 &=& 2.14\times 10^{-7}  \nonumber \\
   \epsilon_3 &=& 2.82 \times 10^{-11}
   \label{eq:E34_eps}
\end{eqnarray}

Table~\ref{tab:E34EmittanceIPAC} summarizes the symplectic eigenvalues and eigenvectors of $J\Sigma$, which is compared to Table~\ref{tab:E34SkeletonigenEmittanceMode} for that of skeleton mode.

\begin{table}[!hbt]
   \centering
	\caption{ Symplectic eigenmode of $J\Sigma$ from Fig.~\ref{fig:E34IPAC26_zoom}.}
   \begin{tabular}{lc}
       \toprule
	     &Symplectic Eigenmods' values \\
         &[$\Delta~x$,$\Delta~p_x$,$\Delta~y$,$\Delta~p_y$,$\Delta~z$,$\Delta~p_z$]$\times10^{-2}$ \\ 
       \midrule
     $Re(v_1)$ &[45.47,11.75,6.34,4.77,-51.13,-68.38] \\
     $Im(v_1)$ & [1.20,0.58,-2.10,-5.15,20.22,0.00]  \\ 
     $Re(v_2)$ & [61.68,0.06,-11.90,-20.20,13.62,2.71]\\
    $Im(v_2)$ & [0.00,20.98,58.76,-27.09,-21.56,-19.02]\\ 
    $Re(v_3)$ & [-87.95,-10.70,14.91,-38.39,-10.65,-9.70]\\
    $Im(v_3)$ & [0.00,-4.59,12.26,-7.60,1.89,-3.64]\\ 
          \bottomrule
   \end{tabular}
   \label{tab:E34EmittanceIPAC}
\end{table}

To examine the completeness of the symplectic modal representation, the 1000-particle distribution was reconstructed cumulatively using the first mode, the first two modes, and all three modes listed in Table XI. Defining the global reconstruction residual as
\begin{equation}
    \eta^{(K)}=\frac{\|X-X^{(K)}\|}{\|X\|}
    \label{eq:AppEta}
\end{equation}
where $X^{(K)}$ denotes the reconstruction using the first K symplectic modes, we obtain
\begin{equation}
\eta^{(1)}=0.683,~\eta^{(1+2)}=0.494,~\eta^{(1+2+3)}=1.86\times10^{-15}.
\label{eq:AppEta2}
\end{equation}
The first two modes therefore capture substantial components of the distribution but do not provide a complete reconstruction. Inclusion of the third symplectic mode reduces the residual to numerical precision, confirming that the complete set of three modes provides a full six-dimensional representation of the 1000-particle distribution.

This result is consistent with the dimensional correspondence discussed in Appendix B: three complex symplectic modes provide the complete six-dimensional real representation, while organizing the phase space into symplectic modal pairs rather than variance-ranked PCA/SVD directions.



The first principal component accounts for approximately $70.5\%$ of the total variance. Its direction is closely aligned with $J\mathrm{Re}(v_1)$, with an absolute normalized inner product of 0.994.

\begin{figure}[!htb]
   \centering
   \includegraphics*[width=.9\columnwidth]{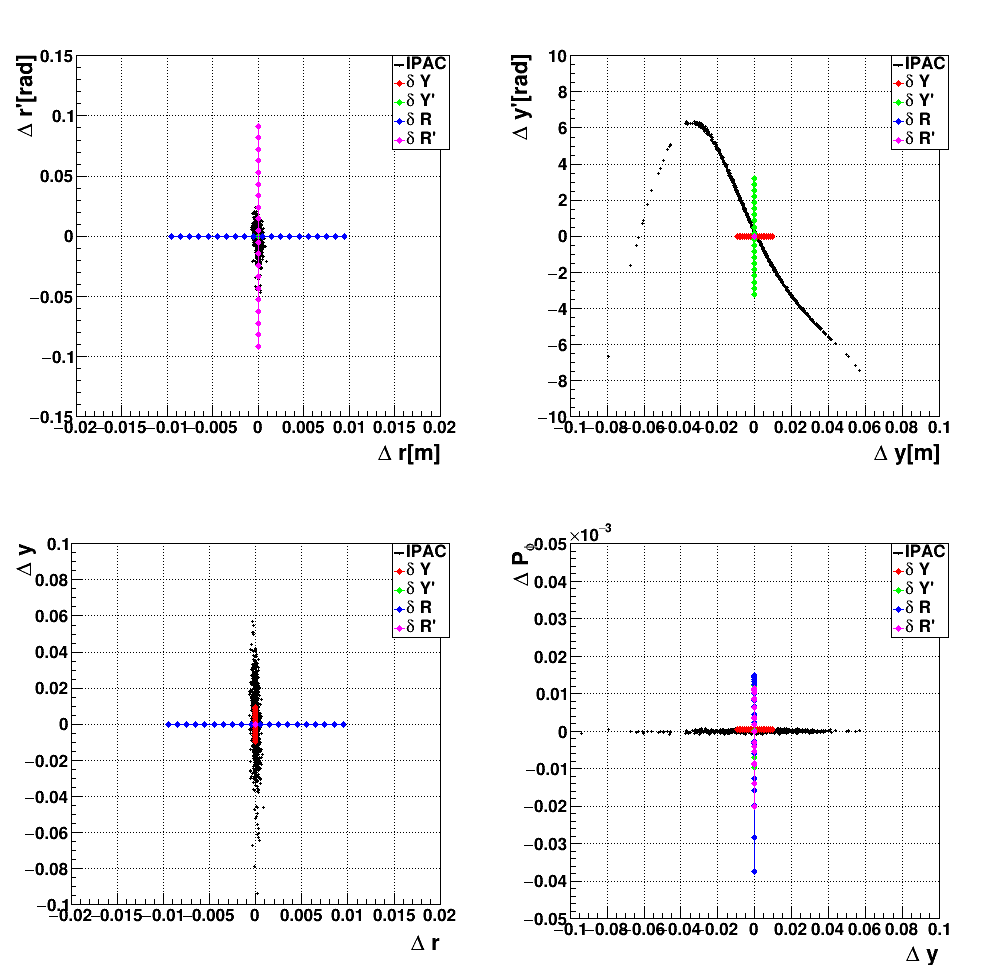}
   \caption{
   Phase-space distributions of the optimized injection beam immediately after the kicker at the end of injection (Y=0). The four panels show the corresponding phase-space projections. The nonlinear transport through the kicker produces significant deformation of the distribution, particularly in the Y–Y'~ plane. The beam is subsequently reverse tracked to the analysis point at Y=0.95 m, where the distributions shown in Fig.~\ref{fig:E34IPAC26_phase} are obtained.}
   \label{fig:E34IPAC26_phase_at0}
\end{figure}

Figure~\ref{fig:E34IPAC26_phase_at0} shows the optimized beam immediately after the kicker, at the end of injection (Y=0). Owing to the nonlinear transport through the kicker, the phase-space distributions, particularly in the Y-Y' plane, exhibit significant distortions. Reverse tracking transports this optimized beam back to the analysis point (Y=0.95 m), where the phase-space distributions shown in Fig.~\ref{fig:E34IPAC26_phase} are obtained. 
The strong deformation of the Y-Y' phase-space distribution is consistent with the observation that the Y- and Y'-perturbed canonical skeletons exhibit nearly identical modal structures.
The covariance matrices used in the following canonical analysis are constructed from these distributions.


Lastly, the first principal component obtained from the PCA of $\Sigma$ is compared with $J\mathrm{Re}(v_1)$ in Table~\ref{tab:E34PCAIPC}. A close alignment is observed, consistent with the correspondence found in the canonical examples presented in the main text.

\begin{table}[!hbt]
   \centering
	\caption{ PCA of $\Sigma$ from Fig.~\ref{fig:E34IPAC26_zoom}.}
   \begin{tabular}{lc}
       \toprule
         &[$\Delta~x$,$\Delta~p_x$,$\Delta~y$,$\Delta~p_y$,$\Delta~z$,$\Delta~p_z$]$\times10^{-2}$ \\
       \midrule
     $u_1$ &     [14.40,-48.00,-1.11,-14.65,-67.23,52.46]\\
    $J\mathrm{Re}(v_1)$ &[11.75,-45.47,4.77,-6.34,-68.38,51.13]\\ 
       \bottomrule
   \end{tabular}
   \label{tab:E34PCAIPC}
\end{table}

\subsection{Canonical Interpretation of the Optimized Injection Beam}
To interpret the optimized injection beam in terms of the perturbative canonical skeletons introduced in Section~\ref{sec:E34skelton}, the beam distribution is projected onto each canonical modal family. This projection provides a quantitative measure of the similarity between individual particles and the corresponding perturbative skeletons.

The projection residual $\eta$ represents the normalized distance between an individual particle and its projection onto a prescribed canonical modal family. This is the same point-wise measure introduced as $\eta_{1}$~in Chapter~\ref{sec:chap3} for the dominant symplectic modal plane. 
Here, the subscript denotes the perturbative canonical family used as the reference; accordingly, $\eta_y$, $\eta_{y'}$, $\eta_{r}$, and $\eta_{r'}$  quantify the particle-wise projection residuals with respect to the corresponding canonical skeletons. Smaller values of $\eta$ indicate that a particle is well represented by the prescribed canonical family, whereas larger values indicate greater deviation from the reference skeleton.

Whereas $C_6$~characterizes the similarity between the dominant symplectic eigenvectors of the optimized beam and the perturbative canonical families, $\eta$ evaluates the correspondence at the particle level. The two quantities therefore provide complementary information: $C_6$ characterizes modal similarity, whereas $\eta$ quantifies how well individual particles are represented by each prescribed canonical family.

The quantitative comparison based on $C_6$~is summarized in Table~\ref{tab:E34C6}.
\begin{table}[!hbt]
   \centering
	\caption{ Comparisons of $C_6$ from Eq.~\ref{eq:c6}.}
   \begin{tabular}{lcccc}
       \toprule
	     skeleton family & $\delta~y$ & $\delta~y'$ & $\delta~r$ & $\delta~r'$\\
       \midrule
	     $C_6 $    &0.773 & 0.765&0.689&0.050\\
       \bottomrule
   \end{tabular}
   \label{tab:E34C6}
\end{table}

To identify which perturbative canonical family most closely represents the dominant mode of the optimized injection beam, the modal similarity $C_6$ is evaluated between its dominant symplectic eigenvector and those obtained from the $\delta y$-, $\delta$ y'-, $\delta$ r-, and $\delta$ r'-perturbed skeletons. Here, the definition of $C_6$ is identical to that introduced in Chapter~\ref{sec:chap3}, while the dominant eigenvector of the optimized beam is used as the reference mode.

\begin{figure}[!htb]
   \centering
   \includegraphics*[width=0.45\columnwidth]{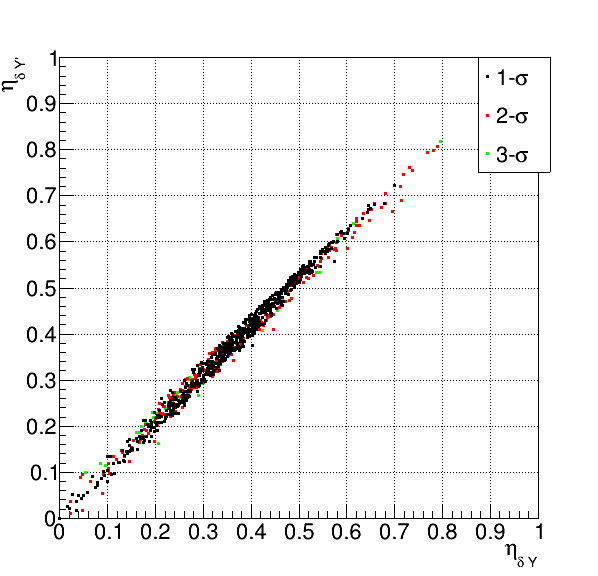}\quad
   \includegraphics*[width=0.45\columnwidth]{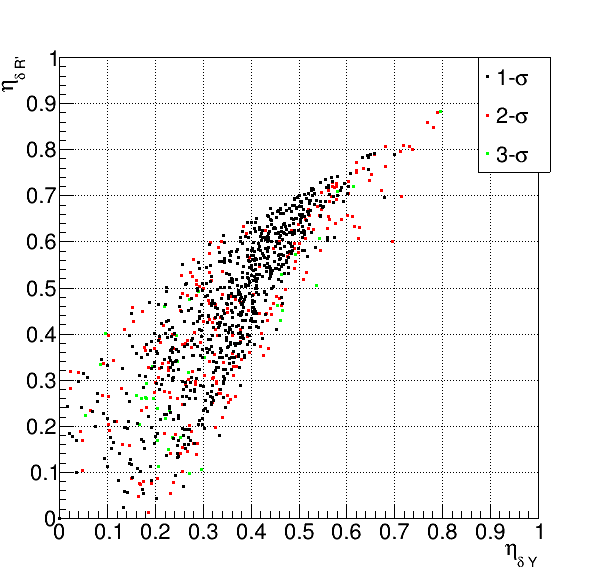}\quad
   \caption{ Left: Comparison of the projection residuals $\eta$ obtained using the Y- and Y'-perturbed canonical skeletons. Each point represents one particle in the optimized injection beam. Black, red, and green points correspond to particles within the 1-$\sigma$, 2-$\sigma$, and 3-$\sigma$ ranges of the canonical angular-momentum distribution, respectively. Right: Comparison of the projection residuals $\eta$ obtained using the Y- and R'-perturbed canonical skeletons. The same particle classification is used as in the left panel.}
   \label{fig:etaPhi_YPYPR}
\end{figure}

To investigate how individual particles are represented by the perturbative canonical skeletons, the projection residual $\eta_{Y}$, $\eta_{Y'}$, $\eta_R$ and $\eta_{R'}$ are evaluated for perturbation families. Left side plot of Fig.~\ref{fig:etaPhi_YPYPR} compares the residuals obtained for the Y- and Y'-perturbed skeletons, while right side plot compares those for the Y- and R'-perturbed skeletons. Each point corresponds to one particle in the optimized injection beam. Black, red, and green points represent particles located within the 1-$\sigma$, 2-$\sigma$, and 3-$\sigma$ ranges of the canonical angular-momentum distribution, respectively.
A strong positive correlation is observed in both comparisons, indicating that particles well represented by the Y-perturbed skeleton are also well represented by the Y'-~and R'-perturbed skeletons.

\begin{figure}[!htb]
   \centering
   \includegraphics*[width=0.45\columnwidth]{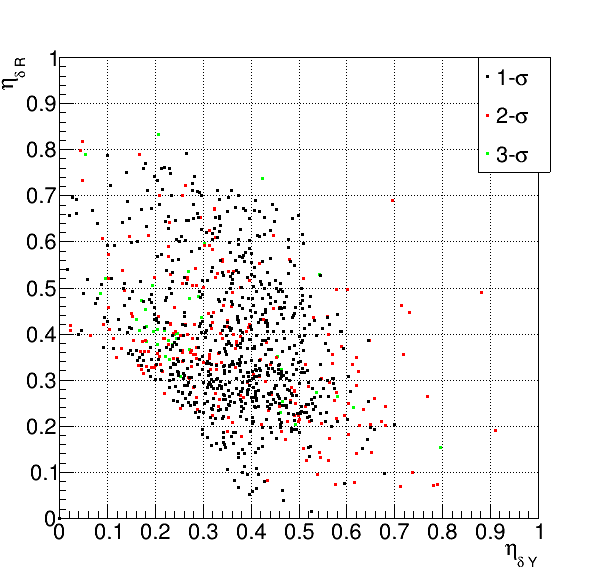}\quad
   \includegraphics*[width=0.45\columnwidth]{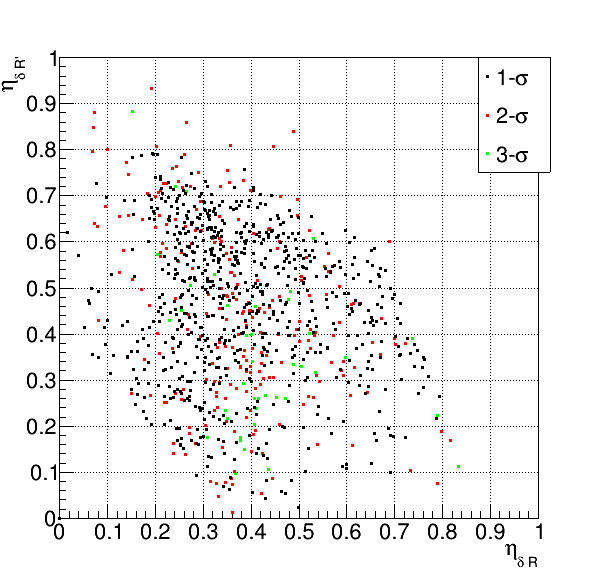}\quad
   \caption{
  Left: Comparison of the projection residuals $\eta$ obtained using the Y- and R-perturbed canonical skeletons. Each point represents one particle in the optimized injection beam. Black, red, and green points correspond to particles within the 1-$\sigma$, 2-$\sigma$, and 3-$\sigma$ ranges of the canonical angular-momentum distribution, respectively. Right: Comparison of the projection residuals $\eta$ obtained using the R- and R'-perturbed canonical skeletons. The same particle classification is used as in the left panel.}
   \label{fig:etaPhi_YRPR}
\end{figure}

Figure~\ref{fig:etaPhi_YRPR} extends the comparison to the Y-R and R-R' perturbation families, allowing the correlations among the four canonical skeletons to be examined systematically.

The projection analysis reveals a strong correlation between the Y- and Y'-perturbed skeletons, indicating that these two perturbation families provide nearly equivalent canonical representations of the optimized beam. In contrast, the R-perturbed skeleton exhibits a substantially weaker correlation, suggesting that it represents a geometrically distinct canonical component.

These results indicate that the optimized injection beam is primarily organized around the Y-type canonical modal family, while the R-type perturbation provides an independent canonical contribution. The present analysis therefore demonstrates how an optimized injection beam can be interpreted in terms of canonical modal families, rather than proposing an alternative beam-design procedure.

The projection analysis further suggests that the statistical formation of optimized injection beam distributions around these canonical skeletons remains an open problem. Clarifying this relationship will be an important subject of future work.


\clearpage
\subsection{Symplectic Eigen Vectors}\label{sec:SymplecticEigen}
This Appendix provides supplementary numerical information on the symplectic eigenmodes discussed in the main text. While the main text focuses primarily on the dominant mode that characterizes each canonical beam family, the tables presented here include the first two symplectic eigenmodes for the $\delta$ y-, $\delta$ y'-, $\delta$~r-, and $\delta$~r'-perturbed skeleton families, together with the corresponding results for the statistically broadened beam distributions. These data are provided to complement the modal characterization in the main text and to allow direct comparison of the dominant and subdominant canonical structures.

\textbf{\textit{$\delta$y-perturbed skeleton}}
The symplectic eigenvalues obtained from the $J\Sigma$ analysis of the $\delta$~y-perturbed skeleton are given in Eq.~\ref{eq:skelton_eps}. 
For the skeleton distribution considered here, these eigenvalues are used primarily to characterize the relative hierarchy of the symplectic modes, rather than as physical beam-emittance measures. 
The eigenvectors corresponding to the first and second symplectic modes are listed in Table~\ref{tab:eigensystemApp_y}.
Table~\ref{tab:PCAIPC_y} further presents the PCA eigenvalues and compares the first principal component with $J\mathrm{Re}(v_1)$.
\begin{eqnarray}
    \epsilon_1&=& 1.91\times 10^{-6} \nonumber\\ 
   \epsilon_2 &=& 2.34\times 10^{-12} \nonumber \\ 
   \epsilon_3 &=& 0.00
   \label{eq:skelton_eps}
\end{eqnarray}

\begin{table}[!hbt]
   \centering
	\caption{ Related vectors obtained from the $J\Sigma$ analysis of Fig.~\ref{fig:perturbAll-03}.
    Despite this increase in the first eigen-emittance, the direction of the first eigenvector is found to remain essentially unchanged from Table~\ref{tab:puturbMode}.}
   \begin{tabular}{l|c}
       \toprule
	     &  Symplectic eigenmode values\\
    $\delta$y & [$\Delta~x$,$\Delta~p_x$,$\Delta~y$,$\Delta~p_y$,$\Delta~z$,$\Delta~p_z$]$\times10^{-2}$ \\
       \midrule
      Re($v_1$) &[ 14.74, -7.43, -31.13, -15.22, -91.81, -9.64]   \\ 
      Im($v_1$) &[ 1.90,  0.15, -0.97, -0.21,  0.00, -0.06 ]\\
       $\|v_1\|$ &[14.87,  7.43,  31.14,  15.22,  91.81,  9.64] \\
       \midrule
     Re($v_2$)& [-31.38, -7.36, -34.63, -12.87,  74.33,  2.71]\\
	   Im($v_2$)&[31.91,  0.01, -30.76, -9.91,  0.00,  1.01]    \\
       $\|v_2\|$ &[ 44.76,  7.43,  46.31,  16.24,  74.33,  2.89] \\
            \bottomrule
   \end{tabular}
   \label{tab:eigensystemApp_y}
\end{table}

\begin{table}[!hbt]
   \centering
	\caption{ PCA of $\Sigma$ from Fig.~\ref{fig:perturbAll-03}.}
   \begin{tabular}{lc}
       \toprule
       $\delta$y  &[$\Delta~x$,$\Delta~p_x$,$\Delta~y$,$\Delta~p_y$,$\Delta~z$,$\Delta~p_z$]$\times10^{-2}$ \\
       \midrule
     $u_1$ &     [-7.43,-14.76,-15.22,31.14,-9.65,91.83]\\
    $J\mathrm{Re}(v_1)$ &[  7.43, -14.74, -15.22,-1.13,  -9.64, 91.81]\\ 
       \bottomrule
   \end{tabular}
   \label{tab:PCAIPC_y}
\end{table} 
The first principal component accounts for approximately $70.5\%$ of the total variance. Its direction is closely aligned with $J\mathrm{Re}(v_1)$, with an absolute normalized inner product of 0.999$\ldots$.

\textbf{\textit{$\delta$~y'-perturbed skeleton}}
The same for the $\delta$~y'-perturbed skeleton are given in Eq.~\ref{eq:skelton_epsPY}. 
The eigenvectors corresponding to the first and second symplectic modes are listed in Table~\ref{tab:eigensystemApp_PY}.
Table~\ref{tab:PCAIPC_PY} further presents the PCA eigenvalues and compares the first principal component with $J\mathrm{Re}(v_1)$.

\begin{eqnarray}
   \epsilon_1 &=&1.28 \times 10^{-5} \nonumber \\ 
   \epsilon_2 &=& 2.65\times 10^{-8} \nonumber \\ 
   \epsilon_3 &=& 7.69\times 10^{-12}
   \label{eq:skelton_epsPY}
\end{eqnarray}

\begin{table}[!hbt]
   \centering
	\caption{ Related vectors obtained from the $J\Sigma$ analysis of Fig.~\ref{fig:perturbAll-03}.
    This table presents the first and second symplectic eigenmodes for the $\delta$~y'-perturbed skeleton. In the main text, only the dominant mode is shown in Table~\ref{tab:puturbMode}.}
   \begin{tabular}{l|c}
       \toprule
	     &  Symplectic eigenmode values\\
    $\delta$y' & [$\Delta~x$,$\Delta~p_x$,$\Delta~y$,$\Delta~p_y$,$\Delta~z$,$\Delta~p_z$]$\times10^{-2}$ \\
       \midrule
      Re($v_1$) &[-14.25,  7.75,  30.82,  14.84,  91.98,  9.69]   \\ 
      Im($v_1$) &[-3.58,  0.06,  0.45, -0.19,  0.00,  0.16  ]\\
       $\|v_1\|$ &[14.69,  7.75,  30.82,  14.84,  91.98,  9.69] \\
       \midrule
     Re($v_2$)& [-18.03,  1.86,  48.00,  21.27,  82.71,  8.10]\\
	   Im($v_2$)&[2.81,  0.05,  0.33,  0.62,  0.00,  0.09]    \\
       $\|v_2\|$ &[ 18.24,  1.86,  48.00,  21.28,  82.71,  8.10] \\
            \bottomrule
   \end{tabular}
   \label{tab:eigensystemApp_PY}
\end{table}

\begin{table}[!hbt]
   \centering
	\caption{ PCA of $\Sigma$ from Fig.~\ref{fig:perturbAll-03}.}
   \begin{tabular}{lc}
       \toprule
       $\delta$y'  &[$\Delta~x$,$\Delta~p_x$,$\Delta~y$,$\Delta~p_y$,$\Delta~z$,$\Delta~p_z$]$\times10^{-2}$ \\
       \midrule
     $u_1$ &     [7.75,-14.29,-14.86,30.88,-9.69,92.02]\\
    $J\mathrm{Re}(v_1)$ &[7.75,-14.25,  -14.84,  30.82,-9.69,91.98  ]\\ 
       \bottomrule
   \end{tabular}
   \label{tab:PCAIPC_PY}
\end{table}

\textbf{\textit{$\delta$~r-perturbed skeleton}}
The same for the $\delta$~r-perturbed skeleton are given in Eq.~\ref{eq:skelton_epsR}. 
The eigenvectors corresponding to the first and second symplectic modes are listed in Table~\ref{tab:eigensystemApp_r}.
Table~\ref{tab:PCAIPC_r} further compares the first PCA eigenvector of $\Sigma$ with $J\mathrm{Re}(v_1)$ derived from the dominant symplectic eigenvector.

\begin{eqnarray}
    \epsilon_1 &=&5.75 \times 10^{-8} \nonumber \\ 
   \epsilon_2 &=& 1.83\times 10^{-11} \nonumber \\ 
   \epsilon_3 &=& 0.00
   \label{eq:skelton_epsR}
\end{eqnarray}

\begin{table}[!hbt]
   \centering
	\caption{ Related vectors obtained from the $J\Sigma$ analysis of~Fig.~\ref{fig:perturbAll-03}.
    This table presents the first and second symplectic eigenmodes for the $\delta$~r-perturbed skeleton. In the main text, only the dominant mode is shown in Table~\ref{tab:puturbMode}.}
   \begin{tabular}{l|c}
       \toprule
	     &  Symplectic eigenmode values\\
    $\delta$~r & [$\Delta~x$,$\Delta~p_x$,$\Delta~y$,$\Delta~p_y$,$\Delta~z$,$\Delta~p_z$]$\times10^{-2}$ \\
       \midrule
      Re($v_1$) &[ -56.25,  20.17, -23.36, 6.32, -74.80, -15.63]   \\ 
      Im($v_1$) &[ -1.88,  0.63, -0.27,  0.27,  0.00, -0.09 ]\\
       $\|v_1\|$ &[ 56.28,  20.18,  23.36,  6.32,  74.80,  15.63] \\
       \midrule
     Re($v_2$)& [  0.79, -0.01,  22.64,  2.88,  85.59, 0.15 ]\\
	   Im($v_2$)&[-40.49,  13.88, -7.88,  6.07, 0.00, -0.749]    \\
       $\|v_2\|$ &[ 40.50,  13.88,  23.97,  6.72,  85.59 ] \\
            \bottomrule
   \end{tabular}
   \label{tab:eigensystemApp_r}
\end{table}

\begin{table}[!hbt]
   \centering
	\caption{ PCA of $\Sigma$ from Fig.~\ref{fig:perturbAll-03}.}
   \begin{tabular}{lc}
       \toprule
       $\delta$~r  &[$\Delta~x$,$\Delta~p_x$,$\Delta~y$,$\Delta~p_y$,$\Delta~z$,$\Delta~p_z$]$\times10^{-2}$ \\
       \midrule
     $u_1$ &     [20.14,56.13,6.30,23.33,-15.66,74.93]\\
    $J\mathrm{Re}(v_1)$ &[20.17,56.25, 6.32,  23.36, -15.63, 74.80   ]\\ 
       \bottomrule
   \end{tabular}
   \label{tab:PCAIPC_r}
\end{table}

\textbf{\textit{$\delta$r'-perturbed skeleton}}
The same for the $\delta$~r'-perturbed skeleton are given in Eq.~\ref{eq:skelton_epsPR}. 
The eigenvectors corresponding to the first and second symplectic modes are listed in Table~\ref{tab:eigensystemApp_PR}.
Table~\ref{tab:PCAIPC_PR} further presents the PCA eigenvalues and compares the first principal component with $J\mathrm{Re}(v_1)$.

\begin{eqnarray}
   \epsilon_1 &=&5.09 \times 10^{-6} \nonumber \\ 
   \epsilon_2 &=& 8.70\times 10^{-11} \nonumber \\ 
   \epsilon_3 &=& 0.0
   \label{eq:skelton_epsPR}
\end{eqnarray}

\begin{table}[!hbt]
   \centering
	\caption{ Related vectors obtained from the $J\Sigma$ analysis of~Fig.~\ref{fig:perturbAll-03}.
    This table presents the first and second symplectic eigenmodes for the $\delta$~r'-perturbed skeleton. In the main text, only the dominant mode is shown in Table~\ref{tab:puturbMode}.}
   \begin{tabular}{l|c}
       \toprule
	     &  Symplectic eigenmode values\\
    $\delta$~r' & [$\Delta~x$,$\Delta~p_x$,$\Delta~y$,$\Delta~p_y$,$\Delta~z$,$\Delta~p_z$]$\times10^{-2}$ \\ 
       \midrule
      Re($v_1$) &[58.24, -2.01,  35.85,  1.99, -72.85,  1.87]   \\ 
      Im($v_1$) &[1.60, -0.24, -1.08, -0.15  0.00, -0.34   ]\\
       $\|v_1\|$ &[58.26, 2.02,  35.86,  1.99,  72.85,  1.90 ] \\
       \midrule
     Re($v_2$)& [-23.89, -9.62, -67.00, -7.33,  68.18, -11.88]\\
	   Im($v_2$)&[ 1.80, -0.33, -1.34, -0.29,  0.00, -0.32]    \\
       $\|v_2\|$ &[ 23.96,  9.63,  67.01,  7.34,  68.18,  11.88] \\
            \bottomrule
   \end{tabular}
   \label{tab:eigensystemApp_PR}
\end{table}

\begin{table}[!hbt]
   \centering
	\caption{ PCA of $\Sigma$ from Fig.~\ref{fig:E34IPAC26_zoom}.}
   \begin{tabular}{lc}
       \toprule
       $\delta$~r'  &[$\Delta~x$,$\Delta~p_x$,$\Delta~y$,$\Delta~p_y$,$\Delta~z$,$\Delta~p_z$]$\times10^{-2}$ \\
       \midrule
     $u_1$ &   
     [-2.01,-58.26,1.99,-35.85,1.87,72.87]\\
    $J\mathrm{Re}(v_1)$ &[ 2.01,58.24,  -1.99,  35.85,  -1.87,  -72.85 ]\\ 
       \bottomrule
   \end{tabular}
   \label{tab:PCAIPC_PR}
\end{table}


\begin{thebibliography}{99}
\bibitem{Iinuma:2016zfu}
H.~Iinuma, H.~Nakayama, K.~Oide, K.~i.~Sasaki, N.~Saito, T.~Mibe and M.~Abe,
Nucl. Instrum. Meth. A \textbf{832}, 51-62 (2016)
doi:10.1016/j.nima.2016.05.126

\bibitem{Iinuma:ipac2025-wepm029}
    H. Iinuma \emph{et al.},
    ``Trajectory design for passing through solenoid magnet fringe field and method for adjusting its strongly X-Y coupled phase space for three-dimensional spiral beam injection'',
    in \emph{Proc. IPAC'25}, Taipei, Taiwan, Jun. 2025, pp. 2020-2023.
    \url{doi:10.18429/JACoW-IPAC2025-WEPM029}

\bibitem{Ogawa:ipac2025-wepm055}
    S. Ogawa \emph{et al.},
    ``Design of beam phase space distribution to realize precise three-dimensional beam injection at J-PARC muon g-2/EDM experiment'',
    in \emph{Proc. IPAC'25}, Taipei, Taiwan, Jun. 2025, pp. 2101-2104.
    \url{doi:10.18429/JACoW-IPAC2025-WEPM055}

\bibitem{PRL2026}
R.~Matsushita, H.~Iinuma, S.~Ohsawa, H.~Nakayama,
K.~Furukawa, S.~Ogawa, N.~Saito, T.~Mibe, and M.~A.~Rehman,
``First Experimental Demonstration of Beam Storage by a Three-Dimensional Spiral Injection Scheme for Ultracompact Storage Rings,''
Phys.\ Rev.\ Lett.\ \textbf{137}, 025002 (2026).
doi:10.1103/8nxx-srgz.


\bibitem{Courant}
E. D. Courant and H. S. Snyder,
"Theory of the Alternating-Gradient Synchrotron,"
Annals of Physics 3, 1–48 (1958).
DOI: 10.1016/0003-4916(58)90012-5

\bibitem{Edwards–Teng}
D. A. Edwards and L. C. Teng,
"Parameterization of Linear Coupled Motion in Periodic Systems,"
IEEE Transactions on Nuclear Science, NS-20, 885–889 (1973).

\bibitem{LebedevBogaczReview}
V. A. Lebedev and S. A. Bogacz,
"Betatron Motion with Coupling of Horizontal and Vertical Degrees of Freedom,"
Journal of Instrumentation 5, P10010 (2010).
DOI: 10.1088/1748-0221/5/10/P10010


~\bibitem{Duftty}
L. Duffy and A. J. Dragt,
"Utilizing the Eigen-Emittance Concept for Bright Electron Beams,"
Advances in Imaging and Electron Physics, 193, 1–44 (2016).
DOI: 10.1016/bs.aiep.2015.11.001


~\bibitem{Groning}
L. Groening, C. Xiao, and M. Chung,
``Particle beam eigenemittances, phase integral, vorticity, and rotations,''
Phys. Rev. Accel. Beams \textbf{24}, 054201 (2021).
DOI: 10.1103/PhysRevAccelBeams.24.054201


~\bibitem{Xu}
W. Xu, P. Piot, J. Ruan, et al.,
"Demonstration of eigen-to-projected emittance mapping for an ionization-cooled beam,"
Physical Review Accelerators and Beams 25, 044001 (2022).
DOI: 10.1103/PhysRevAccelBeams.25.044001



\bibitem{Xiao}
C. Xiao, M. Maier, X. N. Du, P. Gerhard, L. Groening,
S. Mickat, and H. Vormann,
``Rotating system for four-dimensional transverse rms-emittance measurements,''
Phys. Rev. Accel. Beams \textbf{19}, 072802 (2016).
DOI: 10.1103/PhysRevAccelBeams.19.072802


\bibitem{Iinuma:2011zz}
H.~Iinuma, \textquotedblleft{J-PARC muon g-2/EDM}\textquotedblright,
J. Phys. Conf. Ser. \textbf{295}, 012032 (2011)
\url{doi:10.1088/1742-6596/295/1/012032}

\bibitem{BMAG}
  K. L. Brown and T. O. Raubenheimer
"A New Definition of Beam Mismatch and Applications to Linac Beam Transport""(1992)
    



\bibitem{Iinuma:ipac2026-thp5620}
    H. Iinuma \emph{et al.},
    ``Beamline Design and Diagnostics for Strong X–Y Coupled Beams Using a Rotating Quadrupole'',
    presented at IPAC'26, Deauville, France, May 2026, paper THP5620, to be published.


\bibitem{RehmanThesis}
M. A. Rehman, ”A Validation Study on the Novel Three-
Dimensional Spiral Injection Scheme with the Electron Beam
for Muon g – 2/EDM Experiment”, Ph.D. thesis, The Graduate University for Advanced Studies, Sokendai, Japan. \url{http://id.nii.ac.jp/1013/00006023/}

\bibitem{Matsushita:ipac2023-mopa118}
    R. Matsushita \emph{et al.},
    ``Demonstration of three-dimensional spiral injection for the J-PARC muon g-2/EDM experiment'',
    in \emph{Proc. IPAC'23}, Venice, Italy, May 2023, pp. 327-330.
    \url{doi:10.18429/JACoW-IPAC2023-MOPA118}
    

\bibitem{E34PTEP}
M.~Abe~\textit{et al}.,
\textquotedblleft{A new approach for measuring the muon anomalous magnetic moment and electric dipole moment}\textquotedblright,
Progress of Theoretical and Experimental Physics, Volume 2019, Issue 5, May 2019, 053C02.\url{doi:10.1093/ptep/ptz030}


\bibitem{opera} 
Opera, https://www.3ds.com/products/simulia/323opera.

\bibitem{Iinuma:ipac2026-thp5621}
    H. Iinuma \emph{et al.},
    ``Study on evaluation method for transient response of conductor surface to large pulse current'',
    presented at IPAC'26, Deauville, France, May 2026, pp. 4315-4318.

\bibitem{Abe:NIMA890}  
M. Abe, Y. Murata, H. Iinuma, T. Ogitsu, N. Saito, K.
Sasaki, T. Mibe, H. Nakayama, “Design method and
candidate of a magnet for muon g-2/EDM precise
measurement in a cylindrical homogeneous volume”,
Nuclear Inst. and Methods in Physics Research, A:
Volume 890, 11, pp. 51-63, May 2018.
\url{doi:10.1016/j.nima.2018.01.026}
\end{thebibliography}
\end{document}